\documentclass[aps,pre,twocolumn,showpacs,superscriptaddress,groupedaddress,reprint]{revtex4-1}

\usepackage{subfigure}
\usepackage{wrapfig}
\usepackage{textcomp}
\usepackage{inputenc}
\usepackage{enumitem}
\usepackage{graphicx}
\usepackage{amssymb}
\usepackage{amsmath}
\usepackage{bm}
\usepackage{amsmath, amsthm, amssymb}
\usepackage{hyperref}
\usepackage{color}
\usepackage{rotating}
\usepackage{afterpage}
\usepackage{tabulary,booktabs}

\usepackage{amsfonts}
\usepackage{dcolumn}
\usepackage{float}
\usepackage{bm}

\begin{document}
\setcounter{table}{0}

\title{Bootstrapping time correlation functions of molecular dynamics}
\author{Nicolas Desbiens}
\email{nicolas.desbiens@cea.fr}
\affiliation{CEA, DAM, DIF, 91297 Arpajon, France}
\author{Philippe Arnault}
\email{philippe.arnault@cea.fr}
\affiliation{CEA, DAM, DIF, 91297 Arpajon, France}
\author{William Weens}
\affiliation{CEA, DAM, DIF, 91297 Arpajon, France}
\author{Guillaume Perrin}
\affiliation{COSYS, Universit\'e Gustave Eiffel, 77420 Champs-sur-Marne, France}
\author{Vincent Dubois}
\email{vincent-jp.dubois@cea.fr}
\affiliation{CEA, DAM, DIF, 91297 Arpajon, France}

 \date{\today}

\begin{abstract}

Molecular dynamics is often considered as a numerical experiment. The error bars on the results are therefore mandatory, but sometimes difficult to determine and computationally demanding. As a low-cost approach, we describe the application of the bootstrap (BS) method to the quantification of uncertainties pertaining to the time correlation functions. We chose the  autocorrelation functions of velocity and interdiffusion current for a binary ionic mixture as a test bed, and we assessed the merit of the Darken approximation relating both of them. The intrinsic errors related to phase space sampling is investigated comparing the BS method with the reference method of replica. We also study how the BS method can assist in addressing the finite size effects. 

\end{abstract}

\maketitle

\section{Introduction}
Molecular dynamics is one of the most realistic simulation tools available to study the behavior of matter. Since its inception in the 1950s, with the high computing power of the new electronic devices, it has laid down new perspectives for the advancement of science \cite{Battimelli2020}. Nowadays, this method reaches a mature state with still ongoing research and developments.  This is acompanied by an emerging plea to take full advantage of statistical methods to quantify uncertainties of simulation results \cite{Grossfield2009, Romo2011, Nicholls2014, Zhang2015, Pranami2015, Kim2015, Patrone2016, Kim2018, Grossfield2018, Maginn2018}. Such error analysis becomes a critical issue to advance progress in the theoretical framework as well as in the applications of the method. This is a particularly acute issue for transport coefficients in the extreme conditions met in inertial confinement fusion (ICF) experiments \cite{Grabowski2020}, but a proper account of uncertainties could also be beneficial to the modeling of equation of state  \cite{Gaffney2018}.

Actually, the quantification of uncertainties is more numerically demanding for dynamical quantities, like transport coefficients and sound speed, than for static quantities, like pressure and energy. Indeed, the static properties are related to average of the instantaneous values of observables, whereas the dynamical quantities are related to their time correlation functions. This dichotomy is not always the rule when entropy is computed from the velocity correlation function, for instance \cite{Lin2003, Desjarlais2013, Robert2015}. 

The time correlation functions of molecular dynamics simulations incorporate the relaxation of fluctuations around equilibrium, as if they were small external excitations of the system. Their analysis gives access to the preferential modes of excitation of the system such as sound waves \cite{Kryuchkov2019}. The fluctuation - dissipation theorem also allows us to relate some of these functions to the transport coefficients of the hydrodynamic equations \cite{Hansen2006}.

It is computationally demanding to quantify the uncertainty on these time correlation functions, since the estimate based on grouping time sequences by blocks and performing a statistical analysis of the deviations between blocks requires very long duration of the simulation \cite{Bitsanis1987, Chitra1997}. Within this "block averaging" method \cite{Allen2017, Flyvbjerg1989}, the blocks can also be replaced by several simulations, considered as "replica", starting with different initial values to enhance the sampling of phase space. This requirement of long simulation and/or several replicate simulations is even more challenging for costly \textit{ab initio} simulations, where the potential energy surface is recomputed at each time step.

Besides the need to sample thoroughly phase space, the molecular simulations also face an issue of finite-size effects due to the practical limitation on the number of atoms to evolve in the simulation box. This is particularly acute in critical-point conditions \cite{Das2006}. When periodic boundary conditions are used as a remedy to approximate the thermodynamic limit of large number of particles, artifacts incompatible with a fluid behavior may appear \cite{Yeh2004}. Furthermore, spatial lengthscale of correlation can also be squeezed in a  too small simulation box \cite{Petravic2004, Kim2018}. 
To our knowledge, the only way to control the finite-size effect is to perform simulations with larger and larger number of atoms. Since this solution can become prohibitively costly, extrapolation procedures have been proposed with more or less success \cite{Yeh2004, Jamali2018, Sellan2010, Kim2019, Leverant2020, Celebi2020, Dornheim2021}.

As a less expensive alternative to the computationally expensive quantification of uncertainty related to phase space sampling, we investigate in this paper the bootstrap (BS) method \cite{Efron1979, Chernick2008}. It is very simple to implement as a post-processing. With only one single simulation, it provides the poor-man tool to begin with. For \textit{ab initio} simulations, it might be the only path to uncertainties quantification at a reasonable cost. The finite-size effects can be explored when the bootstrap method has provided error bars as a metric. 

The BS method is based on resampling without adding new information. To approximate the sampling distribution of a statistic, one gets "surrogate population" by resampling with replacement from the sample data at hand, creating a large number of synthetic samples known as bootstrap samples. The sample summary (mean value of some observable, for instance) is then computed on each of these BS samples. The histogram of the set of computed values is the bootstrap distribution of the statistic. 
This approach to uncertainty quantification makes no assumption about the particular statistic of the observable of interest, apart for independence of the samples. 

Since the BS method assumes that the sample data are uncorrelated, it needs some adaptation when applied to time series, such as the time correlation functions of molecular dynamics. 
In this paper, we adapt to the autocorrelation functions (ACF) of molecular dynamics the scheme of moving-block bootstrap \cite{Carlstein1986, Kunsch1989}, where the resampling is performed on blocks of consecutive data, in order to preserve the underlying structure of correlation. 

BS technics is rarely applied in the context of molecular dynamics for the properties of systems in thermodynamic equilibrium. Only recently, its efficiency was probed by Crysup and Shanbhag \cite{Crysup2019} for the self-diffusion coefficient as obtained from the mean square displacement, according to Einstein's relation \cite{Hansen2006}. 
 To our knowledge, a bootstrap analysis of time correlation functions has not yet been performed. The uncertainty pertaining to their determination by molecular dynamics is often estimated according to Zwanzig's analysis \cite{Zwanzig1969, Kim2015, Kim2018} under the assumption that the observable follows a Gaussian process.

In this paper, we assess the merit of the BS method at providing with reasonable numerical resources some reliable estimates of error for the interdiffusion current autocorrelation function (JACF) and the velocity autocorrelation function (VACF) giving respectively the mutual diffusion and the self-diffusion coefficients in the Green-Kubo formalism \cite{Hansen2006}. The former is an essential transport property for the hydrodynamic modeling of mixing layers between two different media, with numerous applications in technology \cite{Krishna2019}, ICF \cite{Mackay2020, Viciconte2019}, and Astrophysics \cite{Arnett2000}. The latter is useful in obtaining approximation to the mutual diffusion through the Darken relation \cite{Darken1948, Liu2011}, which relates this coefficient solely to the self-diffusion coefficients of each component in the mixture. This approximation is tested in many recent papers \cite{Krishna2005, Krishna2016, Wolff2018, White2019, Clerouin2020, Rosenberger2020} as a robust estimate.
 
As an illustration of the BS method, we chose to reproduce the results obtained by Hansen, Joly, and McDonald \cite{Hansen1985} in 1985 for a binary ionic mixture (BIM) of ionized hydrogen and helium. Since the Darken relation can be derived by neglecting all cross-correlations between different atoms, there is a relation between the JACF and the VACFs, that Hansen \textit{et al.}  tested in three situations of weak, intermediate, and strong correlations among the ions. These correlations are quantified by only one dimensionless coupling parameter $\Gamma$, which gives the magnitude of the Coulomb potential energy of neighboring particles relative to their kinetic energy, defined later on. Interestingly, the Coulomb potential is the power-law interaction with the longest range exacerbating the possibility for the large spatial correlations not to accommodate within the finite-size of the simulation box.  Nowadays, it is possible to perform many replica simulations in order to investigate how the BS method is efficient at providing error estimates. It is also possible to increase the number of atoms used in Hansen \textit{et al.}'s study to evaluate finite-size effects and the convergence to the thermodynamic limit using BS errors. This investigation can serve as a reference for extrapolating the efficiency of the BS method to more numerically demanding \textit{ab initio} simulations. 

We first recall the basics, the definition of the VACF and the JACF, and the method of replica in Sec.\,\ref{theory}. This serves to introduce the correlation time of ACF and how it is used in the partition of the simulation into uncorrelated sequences. The application of the BS method to the ACFs of molecular dynamics is described in Sec.\,\ref{BSmethod} with specific issues for the average in time and the influence of particles. Our molecular dynamics simulations of the BIM are described in Sec.\,\ref{simus}, and the results of our analysis in Sec.\,\ref{results}.

\section{Theory}
\label{theory}

\subsection{Zwanzig's analysis of ACF}
\label{Zwanzig}

For the sake of completeness, and to introduce the concepts and notations, we recall Zwanzig's analysis \cite{Zwanzig1969, Kim2015, Kim2018} of the statistical uncertainty quantification for ACF under the Gaussian process approximation.

Consider a dynamical variable $a(t)$ whose evolution in time is given by molecular dynamics. In classical statistical mechanics, its time correlation function $C(t)$ is defined as
\begin{equation}
C(t) = \left\langle a(s) \,a(s+t)\right\rangle,
\end{equation}
where the average denoted by $\left\langle\,\right\rangle$ is an ensemble average over an equilibrium distribution of initial system points in phase space. According to Liouville's theorem, $C(t)$ does not depend on the arbitrary time origin $s$ \cite{Hansen2006}.

When using only one molecular simulation, the ensemble average is replaced by a time average under the ergodic assumption
\begin{subequations}
\begin{equation}
C(t) = \lim_{T\to \infty} \,C_T(t),
\end{equation}
\,
\begin{equation}
C_T(t) = \dfrac{1}{T-t_{max}} \int_0^{T-t_{max}} ds ~ a(s) \, a(s+t),
\end{equation}
\end{subequations}

\noindent where $t_{max}$ is the maximum time lag one wants to compute, for which $C(t_{max}) \approx 0$. In the time average, the limiting factor to convergence is the duration $T$ of the simulation. 

When the duration $T$ of a simulation is not long enough to completely sample the phase space, the ACF is estimated with a residual error $\epsilon_T(t)$ given by
\begin{eqnarray}
&\epsilon_T(t) &= C_T(t) - C(t), \\ \nonumber
& &= \dfrac{1}{T} \int_0^T ds \, \left[a(s) \, a(s+t) - \left\langle a(s) \,a(s+t)\right\rangle \right],
\end{eqnarray}
where the invariance of $C(t)$ with the time origin $s$ has been used, and we redefine the duration $(T-t_{max})$ to $T$ for the sake of readability.
The ensemble average of this deviation vanishes, $\left\langle \epsilon_T(t) \right\rangle =0$, but its second moment gives the standard error $\sigma_T(t)$ in the estimation of $C(t)$
\begin{equation}
\sigma_T^2(t) = \left\langle \epsilon_T^2(t) \right\rangle.
\end{equation}

This variance involves computing a fourth-order correlation function of $a(t)$ since
\begin{widetext}
\begin{equation}
\label{sigma4}
\sigma_T^2(t) = \dfrac{1}{T^2} \int_0^T ds_1 \,\int_0^T ds_2 \, \left[\left\langle a(s_1) \, a(s_1+t) \, a(s_2) \, a(s_2+t)\right\rangle - \left\langle a(s_1) \,a(s_1+t)\right\rangle \left\langle a(s_2) \,a(s_2+t)\right\rangle \right].
\end{equation}
\end{widetext}
Zwanzig and Ailawadi \cite{Zwanzig1969} proposed to assume that the observable $a(t)$ follows a Gaussian process to simplify the expression of the standard error $\sigma_T(t)$. Indeed, in this case, a fourth-order correlation is given by second-order correlations according to $
\left\langle a_1 \,a_2 \,a_3 \,a_4\right\rangle = \left\langle a_1 \,a_2\right\rangle \,\left\langle a_3 \,a_4\right\rangle + \left\langle a_1 \,a_3\right\rangle \,\left\langle a_2 \,a_4\right\rangle + \left\langle a_1 \,a_4\right\rangle \,\left\langle a_2 \,a_3\right\rangle
$. Since the second-order correlations are nothing else than the ACFs, $C(t)$, which are independent of the time origin, it is possible to perform one of the two integrations with the change of variable $s = s_2 - s_1$ giving rise to the following approximation to $\sigma_T(t)$, using the subscript ''$Z$'' for Zwanzig,
\begin{equation}
\label{sigmaT}
\sigma_Z^2(t) = \dfrac{2}{T} \int_{0}^{T} ds \,\left[C^2(s) + C(t+s)\,C(t-s) \right],
\end{equation}
where the parity of the integrand has been used. The estimator of the standard error $\sigma_Z(t)$ can be evaluated by using Eq.\,\eqref{sigmaT} with the ensemble average $C(t)$ replaced by the time average $C_T(t)$ obtained in the molecular simulation at hand.

Zwanzig and Ailawadi \cite{Zwanzig1969} defined a mean relaxation timescale $\tau$ by
\begin{equation}
\label{tau}
\tau = 2  \int_{0}^{+\infty} ds ~\dfrac{C^2(s)}{C^2(0)},
\end{equation}
and concluded that the standard error $\sigma_T(t)$ is proportional to $\sqrt{2\tau/T}$. Indeed, for $T$ chosen high enough, it varies from $\sqrt{2 \tau/T}\, C(0)$ at $t=0$ to $\sqrt{\tau/T}\, C(0)$ at $t=T$ according to Eq.\,\eqref{sigmaT}.
The defining  equation \eqref{tau} of $\tau$ is an exact result if the ACF decays exponentially as $\exp(-t/\tau)$. 
 We shall use it in the numerical analysis of molecular dynamics simulations.

Zwanzig and Ailawadi \cite{Zwanzig1969} also evaluated the standard error $\tilde \sigma_T(t)$ of the ACF, $\tilde C_T(t)$, normalized by its value at null time lag $C_T(0)$, with the use of the time average instead of the ensemble average $C(0) = \left\langle a(s) \,a(s) \right\rangle$
\begin{equation}
\tilde C_T(t) = \dfrac{C_T(t)}{C_T(0)}.
\end{equation}
They showed that this standard error $\tilde \sigma_T(t)$ is greatly reduced as compared to $\sigma_T(t)$ for the ACF values at short time range. They proposed the following approximation to  
$\tilde \sigma_T(t)$
\begin{equation}
\tilde \sigma_Z^2(t) \approx \dfrac{2 \tau}{T}\, \left[1 - \dfrac{C(t)}{C(0)} \right]^2.
\end{equation}
In obtaining this result, the main approximation is to replace $C_T(0)$ by $C(0) + \epsilon_T(0)$ in order to Taylor-expand the denominator of $\tilde C_T(t)$ to first order in $\epsilon_T(0)/C(0)$.

One could argue that no gain in accuracy is obtained by the latter estimator, $\tilde C_T(t)$, unless a better estimate of $C(0)$ than $C_T(0)$ is available. Nevertheless, this partition of the uncertainty between $C_T(0)$ and $\tilde C_T(t)$ is useful if an analytical fit of $\tilde C_T(t)$ is required as is often the case to compute transport coefficients from the integral in time of ACFs following the Green-Kubo formulas \cite{Maginn2018}. Indeed, this time integral suffers from propagation and accumulation of errors \cite{Kim2018}, which can be alleviated using $\tilde C_T(t)$ instead of $C(t)$ to find an appropriate fit to the ACF.

\subsection{Uncorrelated sequences}
\label{alpha_def}

The ACF obtained by time average, $C_T(t)$, can be given a more practical expression that emphasizes the role of the correlation time $\tau$, Eq.\,\eqref{tau}, \cite{Maginn2018}
\begin{align}
&C_T(t) = \dfrac{1}{M_T} \sum_{k=1}^{M_T} \, a(s_k) \, a(s_k+t), \\
&s_k = (k-1)~\alpha\,\tau,  \nonumber
\end{align}
where the possibility is open to omit the values of time origin $s$, which are redundant due to correlation. 

The parameter $\alpha$ is the number of correlation times $\tau$ between consecutive time origins, $s_k$ and $s_{k+1}$. It controls how the molecular dynamics trajectory is partitioned into more or less independent sequences. Apart from the time saving afforded by this reduction in time origins to use, it is important to realize that, if the value of $\alpha$ is too low, the redundant time blocks do not bring information, whereas, if it is too large, some information is missing in the time average. We shall see in Sec.\,\ref{BStimes} that the best compromise is given by $\alpha = 1$ when a benchmark is done against the replica method.

\begin{figure}[!t]
\begin{center}
\includegraphics[width=0.85\columnwidth]{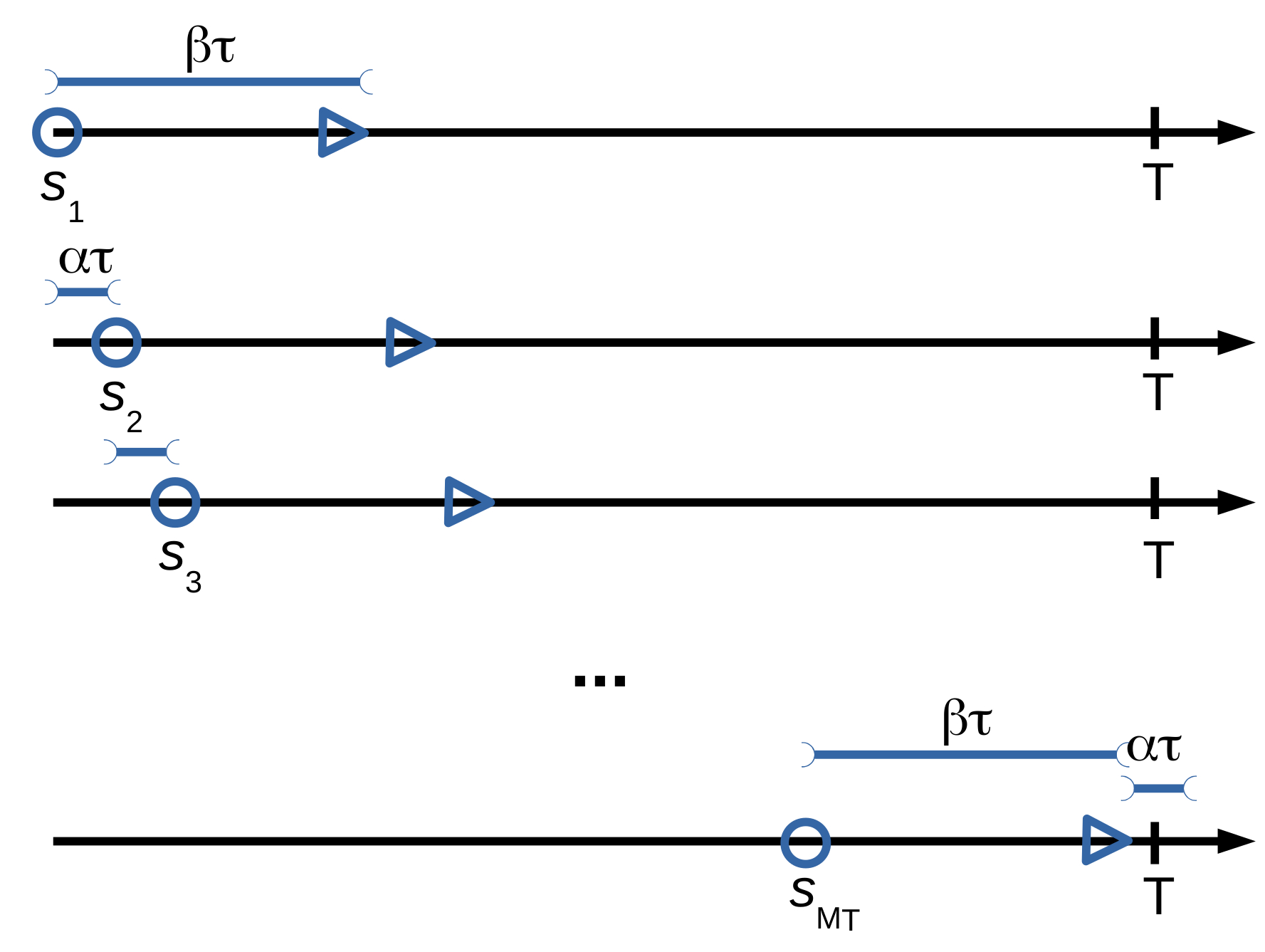}
\caption{Positions of the time sequences of duration $\beta \tau$ used in the calculation of the ACF. The successive time origins $s_k$ are separated by $\alpha \tau$. The last time sequence verifies $s_{M_T} +\beta \tau < T$ and $s_{M_T} +\beta \tau + \alpha \tau > T$, where $T$ is the duration of the simulation.}
\label{scheme_alpha}
\end{center}
\end{figure}

The number $M_T$ of time origins $s_k$ depends on the ratio between the duration $T$ of the trajectory and the lag, equal to $\alpha\,\tau$, between them, but it depends also on the largest time value $t_\text{max}$ for which the ACF $C_T(t)$ is computed, otherwise $(s_{M_T}+t_{max})$ may exceed the total duration $T$. It is practical to introduce a parameter $\beta$ by $t_\text{max} = \beta\,\tau$ . It controls the loss of correlation where $C_T(t_\text{max}) \approx 0$. It should exceed at least $4 \,\tau$. In this study, we adopt the conservative choice of $\beta = 15 \,\tau$. In Fig.\,\ref{scheme_alpha}, a scheme shows the partition of the duration of the simulation into a sequence of time origins, and explains why the value of $M_T$ must fulfill the following constraint
\begin{equation}
\label{nb_s}
\alpha \,(M_T-1) + \beta < T/\tau < \alpha \,M_T + \beta. 
\end{equation}

\subsection{Replica sampling}
\label{replicaMethod}

A reference value of the standard error $\sigma_T(t)$ can be obtained by an ensemble average over $R$ replicate simulations. For each replicate simulation $(r)$, a new value of the ACF is computed  
\begin{align}
&C_T^{(r)}(t) = \dfrac{1}{M_T} \sum_{k=1}^{M_T} \, a^{(r)}(s_k) \, a^{(r)}(s_k+t), \\
&s_k = (k-1)~\alpha\,\tau,  \nonumber
\end{align}
The distribution of replicas is characterized by its mean $C_T^{(R)}(t)$ and its variance $\sigma_R^2(t)$ given by
\begin{subequations}
\begin{equation}
C_T^{(R)}(t) = \dfrac{1}{R}\,\sum_{r=1}^R C_T^{(r)}(t),
\end{equation}
\begin{equation}
\sigma_R^2(t) = \dfrac{1}{R-1}\,\sum_{r=1}^R \left[C_T^{(r)}(t) - C_T^{(R)}(t) \right]^2.
\end{equation}
\end{subequations}
As the number $R$ of replicas increases, the convergence of phase space sampling improves and  $C_T^{(R)}(t)$ tends to the ensemble average $C(t)$, while the standard deviation over replicas $\sigma_R(t)$ divided by $\sqrt{R}$ approaches the standard deviation of the mean $C(t)$. In this limit of large number $R$ of replica, $\sigma_R(t)$ tends to the intrinsic error $\sigma_T(t)$ on each replica simulation due to the finite duration $T$ of the simulation
\begin{equation}
\label{replica_sigma}
\sigma_T(t) = \lim_{R \to \infty}\sigma_R(t).
\end{equation}
To a given value of the parameter $\alpha$, corresponds a partition in time origins, and Eq.\,\eqref{replica_sigma} provides the corresponding error. In Sec.\,\ref{BStimes}, varying $\alpha$ unveils the partition into uncorrelated sequences that leads to the lowest error without redundancy.

\subsection{Velocity and interdiffusion current ACFs}
\label{VACF_JACF}


We shall focus our investigation on two qualitatively different dynamical variables $a(t)$ of a binary mixture: the velocity $\bold v_i^{(m)}(t)$ of each particle $i$ of the component $m$ as an individual property of the particles, and the interdiffusion current $\bold J(t)$ between the two components as a collective property of particles. A time correlation function is defined for each spatial component $v_{i, \chi}^{(m)}$ and $J_{\chi}$ with $\chi$ standing for the Cartesian coordinates $x, y, z$:
\begin{subequations}
\begin{equation}
C_{v_{i, \chi}^{(m)}}(t) = \left\langle v_{i, \chi}^{(m)}(s) \,v_{i, \chi}^{(m)}(s+t)\right\rangle,
\end{equation}
\begin{equation}
C_{J_{\chi}}(t) = \left\langle J_{\chi}(s) \,J_{\chi}(s+t)\right\rangle,
\end{equation}
with 
\begin{equation}
J_{\chi} = \frac{1}{\sqrt N}\,\left[x_2\, J_{1, \chi} - x_1\, J_{2, \chi} \right],
\end{equation}
where
\begin{equation}
\label{Jm}
J_{m, \chi} = \sum_{i=1}^{N_m} v_{i, \chi}^{(m)}.
\end{equation}
\end{subequations}
Here, the binary mixture contains $N_1$ atoms of one species and $N_2$ atoms of the other species. Each species has an atomic concentration $x_m = N_m/N$, with $N = N_1 + N_2$. The factor $1/\sqrt{N}$ ensures that $C_{J_{\chi}}(t)$ does not scale with $N$ \cite{Lebowitz1967}.

In a fluid, the isotropy can be exploited to reduce fluctuations by averaging over the directions \cite{Maginn2018} 
\begin{subequations}
\begin{equation}
C_{\bold v_{i}^{(m)}}(t) = \frac{1}{3} \left[ C_{v_{i, x}^{(m)}}(t) + C_{v_{i, y}^{(m)}}(t) + C_{v_{i, z}^{(m)}}(t)\right],
\end{equation}
\begin{equation}
C_{\bold J}(t) = \frac{1}{3} \left[ C_{J_{x}}(t) + C_{J_{y}}(t) + C_{J_{z}}(t)\right].
\end{equation}
\end{subequations}
The mutual diffusion coefficient is given by the Green-Kubo relation \cite{Hansen2006}
\begin{equation}
D_{12} = \mathcal{J}\,\int_0^\infty dt ~ C_{\bold J}(t),
\end{equation}
where $\mathcal{J}$ is a thermodynamic factor
\begin{equation}
\mathcal{J} = \left.\dfrac{\partial^2 (G/N k_B T)}{\partial^2 x_1}\right|_{P, T, N},
\end{equation}
where $T$ is the temperature ($k_B$, Boltzmann constant), $P$ the pressure, and $G$ the Gibbs free energy of the mixture. This factor expresses the fact that diffusion is driven by the gradient of chemical potential, and that this gradient is driven by the gradient of atomic concentration. Other sources of diffusion include chemical potential gradients induced by temperature and pressure gradients \cite{Landau1992}.

At contrast to the interdiffusion current ACF, the velocity ACF can be further averaged over the particles 
\begin{equation}
\label{vacf_mean}
C_{\bold v^{(m)}}(t) = \frac{1}{N_m} \, \sum_{i=1}^{N_m} C_{\bold v_{i}^{(m)}}(t).
\end{equation}
This VACF is then used to compute a characteristic self-diffusion coefficient $D_m$ of each component, using the Green-Kubo relation \cite{Hansen2006}
\begin{equation}
D_m = \int_0^\infty dt ~ C_{\bold v^{(m)}}(t).
\end{equation}

At this stage, it is worthwhile noting the different way the numbers $N_m$ of particles enters the calculation of the self-diffusion and mutual diffusion coefficients, $D_m$ and $D_{12}$. It is often stated that the VACF, resp. $D_m$, is more accurately determined than the JACF, resp. $D_{12}$, due to the average over $N_m$ particles, with a statistical error reduced by a factor $1/\sqrt {N_m}$. This is only warranted if the trajectories of the particles are uncorrelated. Unfortunately, the interactions among the particles create such correlation, and the gain in accuracy is not obvious as demonstrated by Kim \textit{et al.} \cite{Kim2015}. Moreover, an accurate determination of these coefficients using molecular dynamics simulations presupposes that the thermodynamic limit is reached, \textit{i.e.} that the results do not vary as $N$ increases with $N/V$ constant, where $V$ is the volume of the system. 

Using the BS method, we shall see in Sec.\,\ref{BStimes} and \ref{BSatoms} that the VACF is indeed far more accurately determined than the JACF. We shall also investigate in Sec.\,\ref{BSatoms} how bootstrapping over the time origins and the particles provides a metrics to assess the convergence toward the thermodynamic limit as the number of atoms increases.

\section{Bootstrapping}
\label{BSmethod}

The BS method is applied whenever a complete exploration of the statistic is impractical. It provides information on the statistic from a limited sample of data. In this paper, we apply it to a single molecular dynamics simulation, and try to assess its merit as compared to more expensive approaches, which use many replica simulations with different initial conditions to improve the sampling of phase space, or simulations with larger and larger number of particles to get rid of finite size effects and reach the thermodynamic limit.

Since phase space sampling encompasses the trajectories of every particle, a complete quantification of uncertainties is obtained by a simultaneous BS in time and particles. 

\subsection{Bootstrapping time origins}
\label{BStimeMethod}

Bootstrapping the values of time origin $s_k$ allows to preserve the underlying correlation of the time series $a(t)$ in the spirit of the moving-block BS method \cite{Carlstein1986, Kunsch1989}. A number $B$ of BS samples is generated by drawing with replacement $M_T$ time origins $s_k^{(b)}$ from the original set. Each BS sample $(b)$ is used to compute a new value of the ACF 
\begin{align}
&C_T^{(b)}(t) = \dfrac{1}{M_T} \sum_{k=1}^{M_T} \, a(s_k^{(b)}) \, a(s_k^{(b)}+t), \\
&s_k = (k-1)~\alpha\,\tau.  \nonumber
\end{align}
The BS distribution is characterized by its mean $C_T^{(B)}(t)$ and its variance $\sigma_B^2(t)$ given by
\begin{subequations}
\begin{equation}
\label{BSmean}
C_T^{(B)}(t) = \dfrac{1}{B}\,\sum_{b=1}^B C_T^{(b)}(t),
\end{equation}
\begin{equation}
\label{BSstddev}
\sigma_B^2(t) = \dfrac{1}{B-1}\,\sum_{b=1}^B \left[C_T^{(b)}(t) - C_T^{(B)}(t) \right]^2.
\end{equation}
\end{subequations}
As the number $B$ of BS samples increases, the bias between $C_T^{(B)}(t)$ and $C_T(t)$ is expected to vanish, while the BS standard deviation $\sigma_B(t)$ approaches the standard error $\sigma_T(t)$. We shall see in Sec.\,\ref{BStimes} that this is only warranted for values of $\alpha$ greater than unity that exclude correlated sequences.

\subsection{Bootstrapping particles}
\label{BSatomMethod}

Bootstrapping the particles allows one to estimate the influence on the ACFs of the limited number of particles in the simulation. Doing so, we address several interrelated questions: how fluctuations among the trajectories of particles affect the average over $N_m$ individual VACFs to produce a global VACF, Eq.\,\eqref{vacf_mean}? How they affect the collective variable $\bold J$, Eq.\,\eqref{Jm}, and as a consequence the JACF? Are there significant artifacts due to the periodic boundary conditions? At constant density, is the number of particles enough to get a large enough simulation box to accommodate all the correlation length scales? The brute-force approach to tackle globally these issues is to perform simulations with increasing number of particles and to monitor the convergence of the ACFs. We shall see in Sec.\,\ref{BSatoms} that the BS method provides a metrics to answer these questions.

A number $B$ of BS samples is generated by drawing with replacement $N$ particles from the original set. Each BS sample $(b)$ is used to compute a new value $C^{(b)}(t)$ of an ACF $C(t)$. The BS distribution is characterized by its mean $C^{(B)}(t)$ and its variance $\sigma_B^2(t)$ given by analogs of Eqs.\,\eqref{BSmean} and \eqref{BSstddev}.

In the case of the VACF, $C = C_{\bold v^{(m)}}$, the BS method provides an estimator of the mean, since according to Eq.\,\eqref{vacf_mean} the VACF is the average of $N_m$ individual VACFs of each atom of species $m$. As the number $B$ of BS samples increases, the bias between $C_{\bold v^{(m)}}^{(B)}(t)$ and $C_{\bold v^{(m)}}(t)$ is expected to vanish, while the BS standard deviation $\sigma_B(t)$ is a measure of the fluctuations among the particles present in the simulation box. 

In the case of the JACF, $C = C_{\bold J}$, the BS method is no longer an estimator of the mean, and the expectation of this estimator $C_{\bold J}^{(B)}(t)$ does not coincide with $C_{\bold J}(t)$. In order to interpret its variance $\sigma_B^2(t)$ as a measure of the fluctuations among the particles, it is necessary to renormalize each BS sample by a factor which restore the correct expectation.

To illustrate the difference between individual and collective correlations, and explain how the bootstrap method is adapted in each case, we consider a simplified situation corresponding to a set of $N$ independent random variables $x_i$ of vanishing expectation and equal variance $\sigma^2$,
\begin{subequations}
\begin{equation}
 E(x_i) = 0,
 \end{equation} 
\begin{equation}
 E(x_i x_j) = \sigma^2 \, \delta_{ij}.
 \end{equation}
 \end{subequations}
 This last assumption of independence of the random variables $x_i$ amounts to neglect cross-correlations between different particles. This is the very hypothesis leading to the Darken relation.
 
The correlation function of an individual variable is represented by the summation $S_2$ of the square random variables
\begin{equation}
 S_2 = \sum_{i = 1}^N x_i^2,
 \end{equation} 
whereas the correlation function of a collective variable is represented by the square of the summation $S_1$ of the random variables
\begin{equation}
S_1^2 = \left[\sum_{i = 1}^N x_i\right]^2.
\end{equation}
The expectation values of $S_2$ and $S_1^2$ are the same in this oversimplified example
\begin{equation}
 E(S_2) = E(S_1^2) = N\,\sigma^2.
 \end{equation} 
The goal of the bootstrap method is to provide a simple and economical way to estimate the variance of $S_2$ and $S_1^2$ by sampling with replacement the random variables $x_i$. This gives rise to a distribution of bootstrap samples whose expectation and variance are expected to approximate those of $S_2$ and $S_1^2$.

Selecting bootstrap samples can be represented by a set of independent random variables $I_i$ with values ranging from 1 to $N$, and equal probability
\begin{equation}
P(I_i = j) = \dfrac{1}{N}.
\end{equation}
Here $i$ is the index of the selection of one particle, $i = 1 \dots N$, and $j$ is the index of the picked particle, $j = 1 \dots N$.

When bootstrapping, the quantity $S_2$ is represented by
\begin{subequations}
\begin{equation}
B_2 = \sum_{i = 1}^N x_{I_i}^2,
\end{equation}
and the quantity $S_1^2$ by
\begin{equation}
B_1^2 = \left[\sum_{i = 1}^N  x_{I_i}\right]^2 = \sum_{i = 1}^N  \sum_{j = 1}^N  x_{I_i} \, x_{I_j}.\end{equation}
Since the random variables $I_i$ are independent, it is advantageous to separate the diagonal terms from the others in the above double sum, leading to
\begin{equation}
B_1^2 = \sum_{i = 1}^N x_{I_i}^2 + \sum_{i = 1}^N \,  \sum_{j = 1, j \neq i}^N  x_{I_i} \, x_{I_j}.
\end{equation}

 \end{subequations}

 For the quantity $B_2$, associated with correlation functions of individual variables, the bootstrap method provides an estimation for the error of the mean. In particular, it is assumed that there is no bias between the expectation values of $B_2$ and $S_2$. Indeed, in the present simple case,
 \begin{equation}
 E(B_2) = \sum_{i = 1}^N \, E(x_{I_i}^2),
 \end{equation} 
 with, according to the law of total expectation,
 \begin{equation}
 E(x_{I_i}^2) = \sum_{k = 1}^N E(x_{I_i}^2 | I_i = k)\, P(I_i = k),
 \end{equation}
 where $E(x_{I_i}^2 | I_i = k) = E(x_k^2) = \sigma^2$ is the conditional
 expectation  of $x_{I_i}^2$ knowing that the value of $I_i = k$. Therefore
 \begin{equation}
 E(x_{I_i}^2) = \sum_{k = 1}^N \sigma^2\, P(I_i = k) = \sigma^2,
 \end{equation}
 and 
 \begin{equation}
 E(B_2) = N \sigma^2 = E(S_2).
 \end{equation}
 
 The situation is different for the quantity $B_1^2$, associated with the correlation functions of collective variables, where the expectation value $E(B_1^2)$ is  different from $E(S_1^2)$.  
 A correction to this bias involves scaling each bootstrap sample by the following factor
 \begin{equation}
  f_B = \dfrac{E(S_1^2)}{E(B_1^2)},
  \end{equation}  
  with
 \begin{equation}
 E(B_1^2) = \sum_{i = 1}^N E(x_{I_i}^2) + \sum_{i = 1}^N \,  \sum_{j = 1, j \neq i}^N  E(x_{I_i} \, x_{I_j}).
 \end{equation}
 The first term on the right hand side is equal to  $E(B_2)$. To compute the second term, we apply again the law of total expectation
 \begin{eqnarray}
 E(x_{I_i} \, x_{I_j}) = \sum_{k = 1}^N \sum_{l = 1}^N E(x_{I_i} x_{I_j} | I_i = k, I_j = l)  \nonumber \\
 \times  P(I_i = k) P(I_j = l),
 \end{eqnarray}
 using the independence of the random variables $I_i$. Again, the conditional expectation can then be evaluated
 \begin{equation}
 E(x_{I_i} x_{I_j} | I_i = k, I_j = l) = E(x_k x_l) = \sigma^2 \delta_{kl},
 \end{equation}
 leading to
 \begin{equation}
 E(x_{I_i} \, x_{I_j}) = \sum_{k = 1}^N  \sigma^2  P(I_i = k) P(I_j = k) = \dfrac{\sigma^2}{N},
 \end{equation}
and the final expression
\begin{equation}
E(B_1^2) = (2 N - 1)\, \sigma^2.
\end{equation}
The scaling factor $f_B = N / (2 N - 1)$ tends to the value of $1/2$ as $N$ increases.

The calculations corresponding to the interdiffusion current ACF follow the same line of arguments, leading to a scaling factor 
\begin{equation}
f_B = \dfrac{x_2^2 N_1 \sigma_1^2 + x_1^2 N_2 \sigma_2^2}{x_2^2 (2 N_1-1) \sigma_1^2 + x_1^2 (2 N_2-1) \sigma_2^2},
\end{equation}
which also tends to $1/2$ as $N_1$ and $N_2$ increase.

Although the correlations between the atoms in the simulation are more intricate than in the preceding derivation, the ratio between $C_{\bold J}^{(B)}(t)$ (computed without the renormalization) and $C_{\bold J}(t)$ leads to the same value of $1/2$.

\section{BIM simulations}
\label{simus}

As a test bed to the BS method, we chose a binary mixture of ions in Coulomb interaction. 

\subsection{BIM system}
\label{BIM}

The binary ionic mixture (BIM) model is an extension for mixtures of the one component plasma (OCP) model. The OCP is the simplest model of a Coulomb system, that consists of a single species of ions immersed in a neutralizing background of electrons \cite{Hansen1973, Hansen1975, Baus1980, Daligault2006, Clerouin2016}.  This corresponds to a limiting case of real matter under the extreme conditions of white dwarfs interiors \cite{Koester1990}, for instance, where the electrons are fully degenerate. Although this model is an over-simplification when electrons pile up around ions, an effective OCP can still be defined to serve as a prototype  similar to the hard-sphere model in the theory of simple liquids \cite{Clerouin2016}. Since its properties are either analytical or tabulated, it is used as a practical representation of Coulomb coupling in many situations encountered in hot dense plasmas. Its static and dynamical properties depend on only one dimensionless parameter, the Coulomb coupling parameter $\Gamma$
\begin{equation}
\Gamma = \dfrac{Q^2}{k_B T \,a},
\end{equation}
in atomic units, where $a = (4 \pi n/ 3)^{-1/3}$ is the Wigner-Seitz radius, $n$ is the ionic density, $Q$ is the ionization, and $T$ is here the temperature. As $\Gamma$ increases, the OCP transits continuously from a nearly collisionless, gaseous regime for $\Gamma \ll 1$ to an increasingly correlated, liquidlike regime when $\Gamma \gtrsim 1$, up to the Wigner crystallization near $\Gamma = 175$.

The modeling of the BIM introduces three additional parameters: the atomic concentration $x_1$ ($x_2 = 1 - x_1$), the mass ratio $M_2 / M_1$, and the charge ratio $Q_2 / Q_1$.

There is a practical OCP system of units. Using the Wigner-Seitz radius $a$ as the length scale, a natural time scale is the inverse of the plasma frequency $\omega_P$, defined in atomic units by
\begin{equation}
\omega_P^2 = \dfrac{4 \pi n \,Q^2}{M},
\end{equation}
where $M$ is the mass of ions. Indeed, Newton's equations of motion are then independent of the mass $M$ and the charge $Q$. Moreover, the density $n = 3/4 \pi$ is constant and the temperature $T = 1/3 \Gamma$ is a measure of the coupling. In these OCP units, every properties depend only on the Coulomb coupling parameter $\Gamma$ (see Appendix \ref{appOCP} for details). 

For the BIM of hydrogen (H) and helium (He), we adopted the plasma frequency $\omega_P$ computed with the mass $M_1$ and the charge $Q_1 = 1$ of H, leading to Newton's equations of motion parameterized by $\Gamma$, $M_2 / M_1 = 4$, and $Q_2 / Q_1 = 2$.

\subsection{Molecular dynamics}
\label{MD}

The simulations comprise $N_1$ H ions and $N_2$ He ions in a cubic box periodically replicated in all directions. The initial particle positions are on a body-centered cubic lattice. Their initial velocities are assigned randomly from a Maxwell-Boltzmann distribution at the desired temperature. 

In a microcanonical (NVE) simulation, at constant energy $E$, the particle trajectories are determined by solving Newton's equations of motion with the velocity Verlet integrator \cite{Frenkel2002}. Since the range of the Coulomb interaction is larger than the simulation box, the force on each ion needs to include its interaction with the ions in the simulation box and in the periodically replicated cells. To this end, it is calculated using the Ewald summation technique \cite{Allen2017}. 

Two thermostats are used: a simple velocity scaling is first performed at every time step of the equilibration phase to maintain the desired temperature until the spatial structure of the ions reaches equilibrium configurations; then, the simulation transitions to the production phase in either NVE ensemble or the gaussian isokinetic ensemble (NVK), at constant kinetic energy $K$ \cite{Evans1983, Evans1983a, Evans1984, Minary2003}. The NVK thermostat is useful to compare the results with Hansen \textit{et al.}'s microcanonical simulations \cite{Hansen1985} at given coupling parameter $\Gamma$, \textit{i.e.} at given temperature $T$ in OCP units (see Appendix\,\ref{appOCP}). It is also useful when comparing simulations with different numbers of atoms to avoid fluctuations of temperature (see Sec.\,\ref{BSatoms}). We adapted the velocity Verlet algorithm to the NVK ensemble using an even simpler discretization than Zhang's one \cite{Zhang1997}, which is based on the leap-frog algorithm (Appendix \ref{appNVK}).

Hansen \textit{et al.} \cite{Hansen1985} presented simulations of equimolar mixture of H and He at three values of the coupling parameter $\Gamma$ equal to 0.4, 4, and 40, spanning a large range of correlations from weak to strong couplings. We tried to keep their conditions of simulation with a number of $250$ atoms but the NVK thermostat needed smaller time steps $\Delta t$ listed in Table \ref{table1} with the duration of the equilibration phase, $T_\text{eq}$, and of the production phase, $T_\text{prod}$. The time step $\Delta t_\text{ACF}$ used to compute the ACFs is closer to Hansen's values.

The replica simulations, presented in Sec.\,\ref{replica}, were performed keeping the number of $250$ atoms with different random seeds to generate the initial velocities. Some replica simulations were also conducted at larger number of atoms, corresponding to the evaluation of finite size effects presented in Sec.\,\ref{BSatoms}, where the number of atoms was increased to: 432, 1024, 2000, and 3456. These larger simulations used the same time steps as at 250 atoms.

\begin{table}[b!]
\caption{Conditions of molecular dynamic simulations: $\Gamma$ is the coupling parameter, $\Delta t$ the time step in unit of inverse plasma frequency $\omega_P$, $T_\text{eq}$ the duration of the equilibration phase, $T_\text{prod}$ the duration of the production phase, and $\Delta t_\text{ACF}$ the time step used to compute the ACFs.}
\begin{center}
\begin{tabular}{c c c c c c c c c c}
$\Gamma$	 & & $\omega_P \Delta t$	& & & $T_\text{eq}/\Delta t$ & & $T_\text{prod}/\Delta t$ & & $\Delta t_\text{ACF}/\Delta t $ \\
\hline
0.397	& & 0.02   & & & 5000 & & 15000 & & 6  \\
3.992	& & 0.025  & & & 6000 & & 30000 & & 6  \\
39.738	& & 0.05   & & & 3000 & & 15000 & & 3	 \\
\hline
\end{tabular}
\end{center}
\label{table1}
\end{table}%

\section{Results}
\label{results}

We start the presentation of results with the reference method of replica. The error estimations from replica is then compared with what can be obtained from only one simulation using the BS method. The merit of the BS method is further assessed as concerns the finite size effects. Finally, as an illustration of the use of the BS approach, we reexamine Hansen \textit{et al.}'s comparison \cite{Hansen1985} between the ACF of the interdiffusion current and the Darken relation involving the VACFs of each species.

\subsection{Replica}
\label{replica}

We performed 200 replica simulations for each value of the coupling parameter $\Gamma$ to obtain an estimate of the intrinsic error $\sigma_r$ of an ACF determination due to the finite duration $T$ of a simulation, according to Eq.\,\eqref{replica_sigma}. In agreement with Zwanzig's analysis presented in Sec.\,\ref{Zwanzig}, the error is almost constant for all time lags, within a factor of 2. It suffices only to monitor its mean value.  
Whatever the ACF and the coupling value, by 50 replicas, the errors begin to converge and they no longer vary after 100 replicas. The case of weak coupling, at $\Gamma = 0.4$, is the slowest to converge.

\begin{figure}[]
\begin{center}
\includegraphics[width=0.50\textwidth]{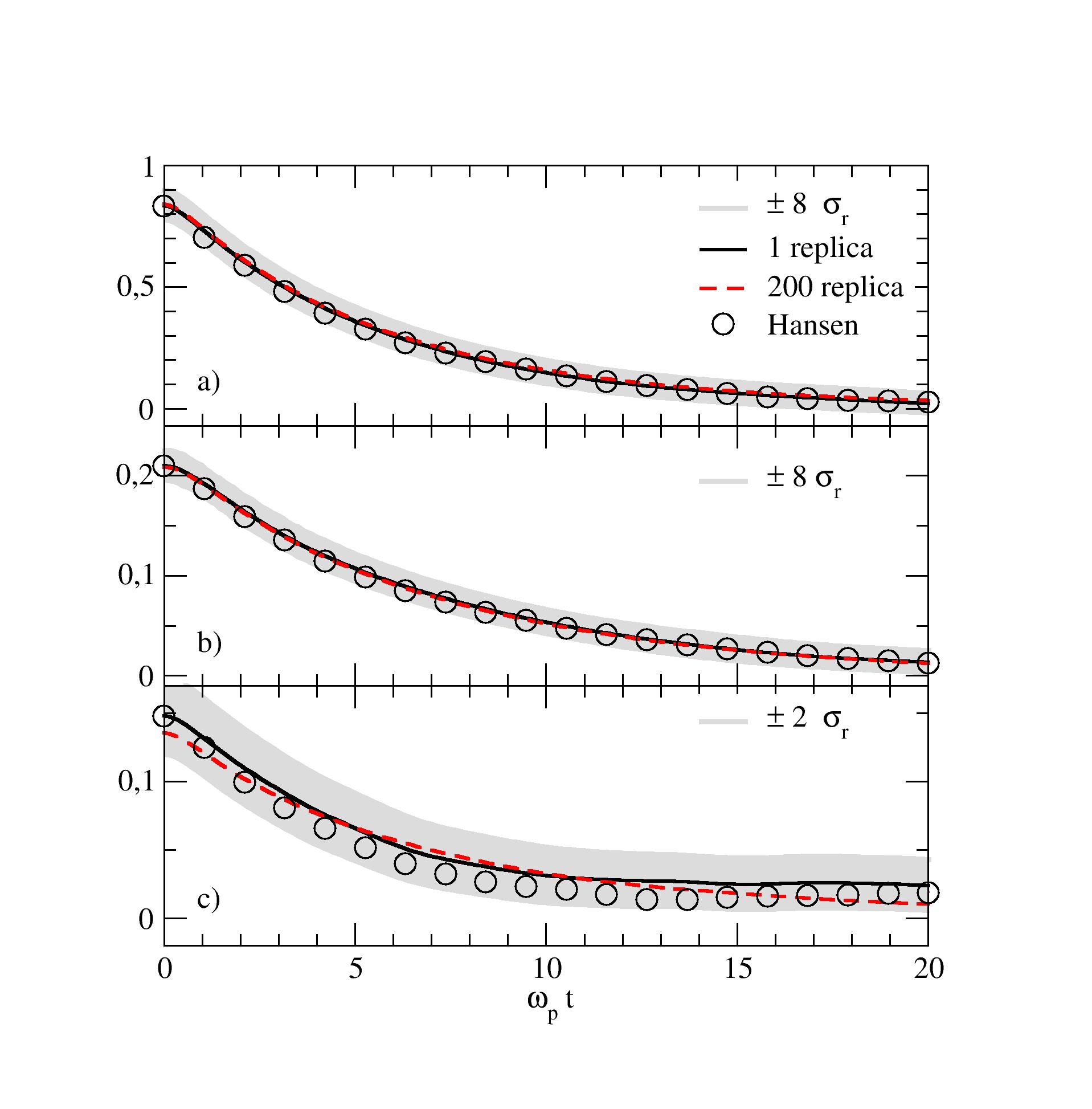}
\caption{ ACFs at $\Gamma = 0.4$ as functions of time lag $t$, in unit of inverse plasma frequency $\omega_P$, with error bars estimated using replica standard deviation $\sigma_r$: a) VACF of H, b) VACF of He, c) JACF. All simulations use 250 atoms. Solid black lines with error bars represent only one replica, dashed red lines the average over 200 replicas, open circles Hansen \textit{et al.}'s data \cite{Hansen1985}.}
\label{ACF_04_replica}
\end{center}
\end{figure}

\begin{figure}[]
\begin{center}
\includegraphics[width=0.50\textwidth]{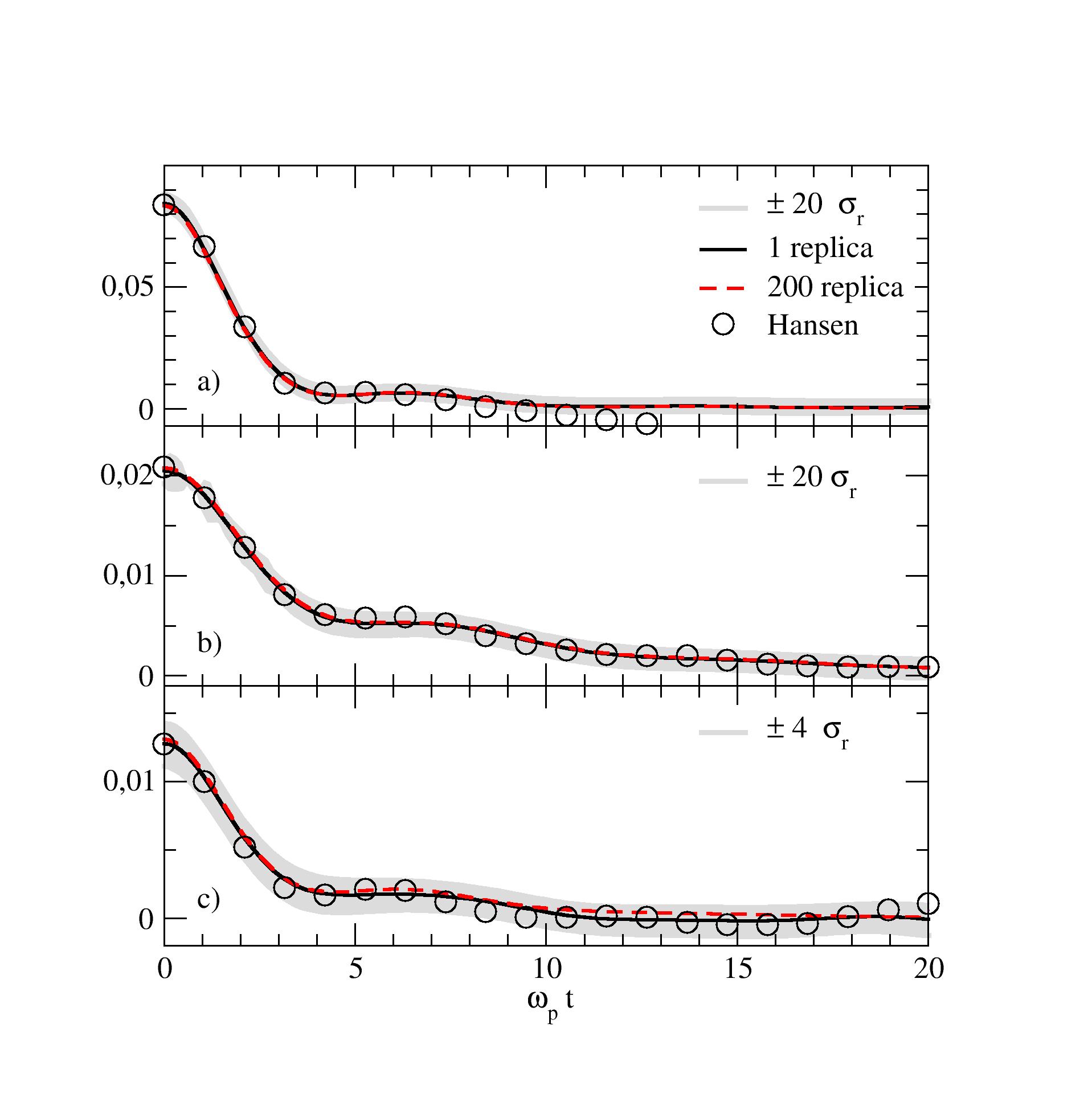}
\caption{ACFs at $\Gamma = 4$ as functions of time lag $t$, in unit of inverse plasma frequency $\omega_P$, with error bars estimated using replica standard deviation $\sigma_r$: same conventions as in Fig.\,\ref{ACF_04_replica}.}
\label{ACF_4_replica}
\end{center}
\end{figure}

\begin{figure}[]
\begin{center}
\includegraphics[width=0.50\textwidth]{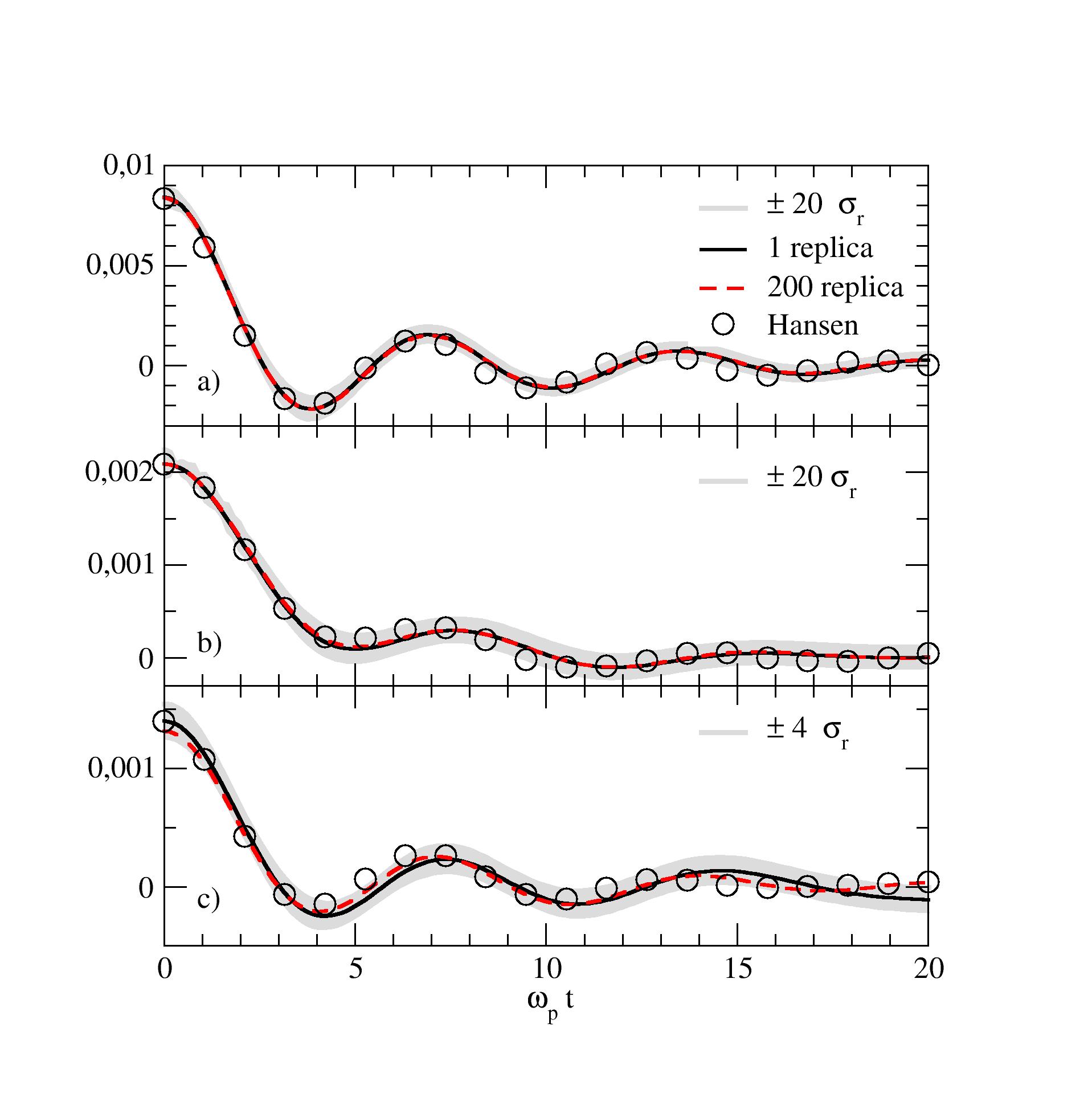}
\caption{ACFs at $\Gamma = 40$ as functions of time lag $t$, in unit of inverse plasma frequency $\omega_P$, with error bars estimated using replica standard deviation $\sigma_r$: same conventions as in Fig.\,\ref{ACF_04_replica}.}
\label{ACF_40_replica}
\end{center}
\end{figure}

The VACFs of H and He, and the JACF of the mixture are represented in Figs.\,\ref{ACF_04_replica} to \ref{ACF_40_replica} for one replica with the error bars derived from 200 replica. Since we are primary interested in the intrinsic errors, the error bars comprise sometimes a very large multiple of the standard deviation $\sigma_r$ of Eq.\,\eqref{replica_sigma}, for the three cases of coupling at $\Gamma$ = 0.4, 4, and 40. This renders visible the errors when they are too small, here and in the following, even though there is no statistical sense in providing error bars larger than 2-4\,$\sigma$.

At weak coupling ($\Gamma = 0.4$), the ACFs decrease with time lag monotonously with a somewhat larger correlation times than at stronger coupling (see Table\,\ref{table2}). Oscillations of the ACFs in time lag appear at intermediate coupling ($\Gamma = 4$) and develop into features of anticorrelation, where the ACFs exhibit negative values, at strong coupling ($\Gamma = 40)$. These anticorrelations originate from a caging effect when a particle bounces against its nearest neighbors. We also plotted Hansen \textit{et al.}'s data \cite{Hansen1985}, which are reproduced with small but significant deviations, that we found difficult to interpret in absence of error bars in their paper. The averages over 200 replica are also plotted. Overall, they lie within 2\,$\sigma_r$ from the results of only one replica. Comparing the error bars, it is evident that the VACFs are determined with higher accuracy than the JACF.

The Table\,\ref{table2} gathers the mean errors $\sigma_r$ obtained with replica. As expected, the ratio between the errors on the VACFs, as ACF of individual observables, and the ones on the JACF, as ACF of collective observable, is close to $1/\sqrt{N_i} \sim 1/10$, where $N_i = 125$ is the number of H or He atoms. The errors on ACFs are larger at $\Gamma = 0.4$ where the correlation time is longer. Zwanzig's Gaussian process approximation, $\sigma_Z$, is included. It overestimates the replica results by around a factor of 2. The mean errors $\sigma_t$ obtained with BS on time origins are included anticipating the results of Sec.\,\ref{BStimes}.

\begin{table*}[t!]
\begin{center}
\caption{Comparison between the estimates of the mean  intrinsic errors obtained with the replica method, $\sigma_r$, the BS on time origins, $\sigma_t$ for $\alpha = 1$, the BS on time origins and atoms, $\sigma_{ta}$, and Zwanzig's Gaussian process approximation, $\sigma_Z$, for the ACFs, $C(t)$, of velocities for species H and He, and the JACF of the mixture. $\Gamma$ is the coupling parameter. $\tau$ is Zwanzig's correlation time, Eq.\,\eqref{tau}, in unit of inverse plasma frequency $\omega_P$. In the case of the VACFs, $\sigma_Z = \sqrt{2 \tau/ N_i T} \, C(0)$ whereas in the case of the JACF $\sigma_Z = \sqrt{2 \tau/ T} \, C(0)$, where $T$ is the duration of the simulation. All simulations of the equimolar mixture used 250 atoms ($N_i = 125$).}
\label{table2}
\begin{tabular}{c c c c c c c c c c c c c c c c c c c c c c c}
$C(t)$ & & $\Gamma$	 & & & $\omega_P \tau$	& & & $C(0)$	& & & $\sigma_r/C(0)$ & & &$\sigma_t/C(0)$ & & &$\sigma_{ta}/C(0)$ & & & $\sigma_Z/C(0)$ \\
 & & 	 & & & 	& & & 	& & & (\%) & & &(\%)  & & &(\%) & & & (\%) \\
\hline
        & & 0.397   & & & 6.1   & & & 0.833    & & & 1.2 & & & 0.8 & & & 1.5 & & & 1.8  \\
VACF H  & & 3.992   & & & 2.7   & & & 0.0843   & & & 0.3 & & & 0.3 & & & 0.6 & & & 0.7  \\
        & & 39.738  & & & 2.7   & & & 0.0084   & & & 0.4 & & & 0.3 & & & 0.6 & & & 0.7  \\
\hline
        & & 0.397   & & & 7.7   & & & 0.210    & & & 1.3 & & & 1.0 & & & 1.9 & & & 2.0  \\
VACF He & & 3.992   & & & 4.8   & & & 0.0205   & & & 0.5 & & & 0.4 & & & 0.8 & & & 1.0  \\
        & & 39.738  & & & 3.6   & & & 0.0021   & & & 0.5 & & & 0.4 & & & 0.7 & & & 0.8  \\
\hline
        & & 0.397   & & & 7.1   & & & 0.148    & & & 13. & & & 11. & & & 15. & & & 21.  \\
JACF    & & 3.992   & & & 3.3   & & & 0.0128   & & & 4.5 & & & 3.6 & & & 5.7 & & & 9.3  \\
        & & 39.738  & & & 3.1   & & & 0.0014   & & & 3.8 & & & 3.8 & & & 5.1 & & & 9.2  \\ 
\hline
\end{tabular}
\end{center}
\end{table*}%
 

\noindent  The latter are close to the replica ones. The mean errors $\sigma_{ta}$ obtained with BS on time origins and atoms are also included anticipating the results of Sec.\,\ref{BSatoms}. Bootstrapping also the atoms leads to mean errors larger by a factor between 1.3 and 2. 

\subsection{BS on time origins}
\label{BStimes}

The BS method is applied here as a post-processing of the simulation where the ACFs are recomputed from different sets of time origins, drawn with replacement from the origin set. 
The convergence of the error estimate on the VACFs and the JACF is rapid.
 Whatever the ACF and the coupling value, by 20 samples, the errors begin to converge and they no longer vary after 40 samples. The case of weak coupling, at $\Gamma = 0.4$, is the slowest to converge.

As the BS method assumes that each time origin contributing to the average ACF is independent of the others, the resulting error estimate depends on the choice of the parameter $\alpha$ which controls the partition of the simulation into uncorrelated sequences (Sec.\,\ref{alpha_def}). This parameter represents the number of Zwanzig's correlation time $\tau$ between consecutive time origins. The higher it is, the less time origins are used to compute the ACF, with a possible loss of information. The lower it is, the more time origins are used with possible redundancies.

\begin{figure}[t!]
\begin{center}
\includegraphics[width=.5\textwidth]{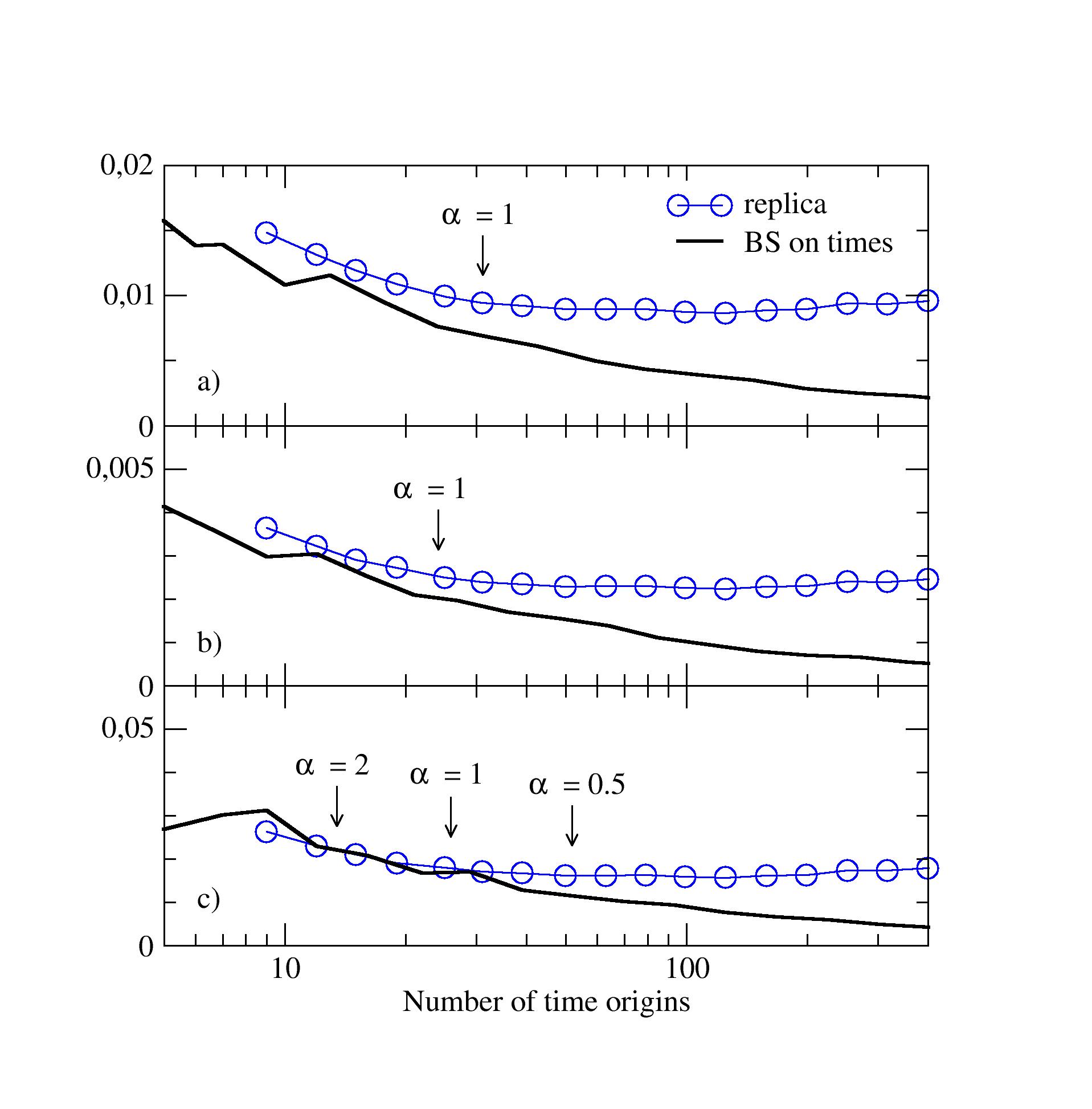}
\caption{ Comparison at $\Gamma = 0.4$ between the BS estimates and the replica results for the mean intrinsic errors on ACF as functions of the number of time origins which depends on the value of $\alpha$ (the number of correlation time $\tau$ between successive time origins): a) VACF of H, b) VACF of He, c) JACF. All simulations use 250 atoms. Solid black lines represent the BS estimates using only one replica, solid blue lines with open circles the results using 200 replicas. }
\label{ACF_04_BS}
\end{center}
\end{figure}
\begin{figure}[t!]
\begin{center}
\includegraphics[width=.5\textwidth]{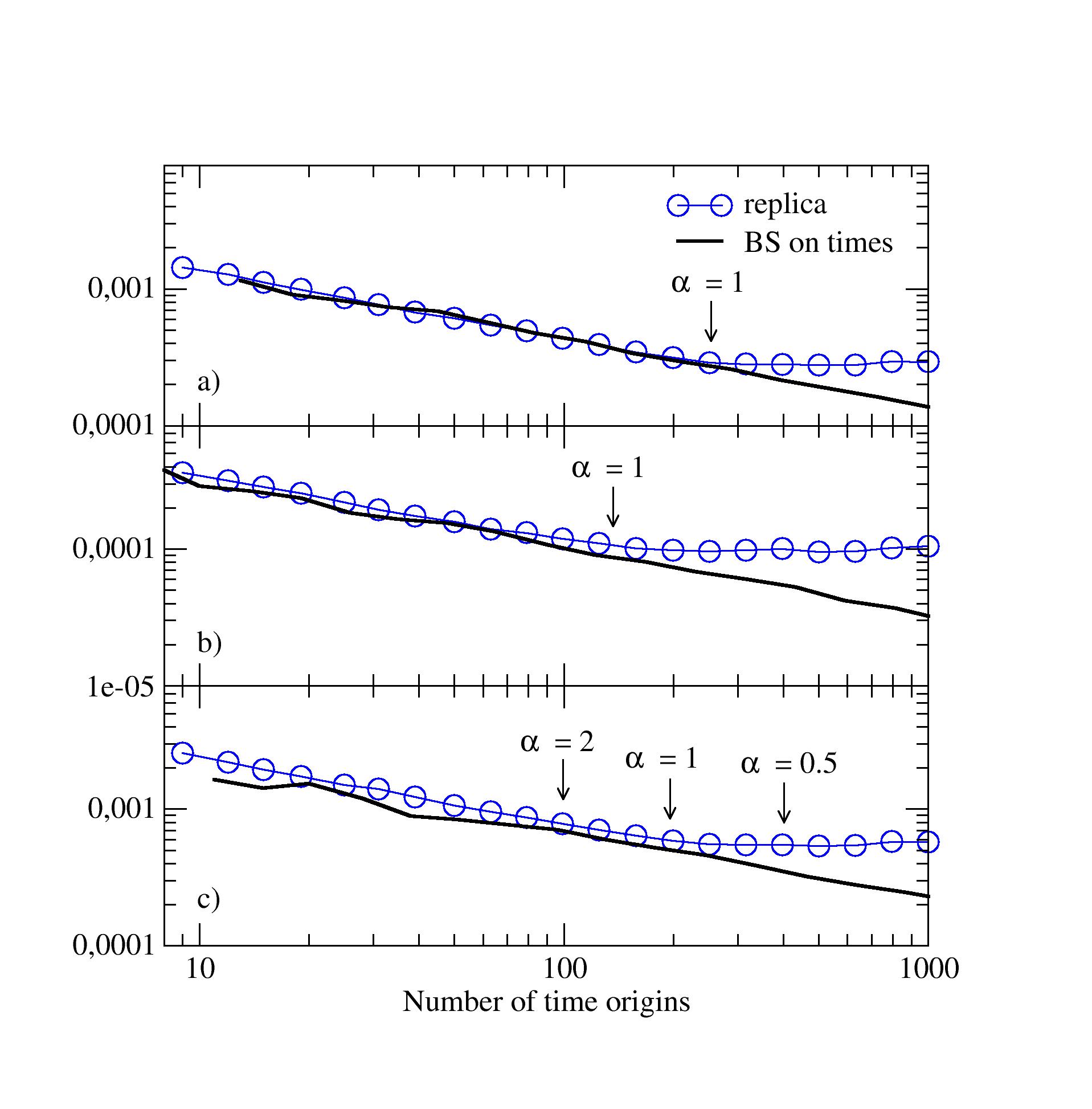}
\caption{Comparison at $\Gamma = 4$ between the BS estimates and the replica results for the mean intrinsic errors on ACF as functions of the number of time origins which depends on the value of $\alpha$: same conventions as in Fig.\,\ref{ACF_04_BS}.}
\label{ACF_4_BS}
\end{center}
\end{figure}
\begin{figure}[h!]
\begin{center}
\includegraphics[width=.5\textwidth]{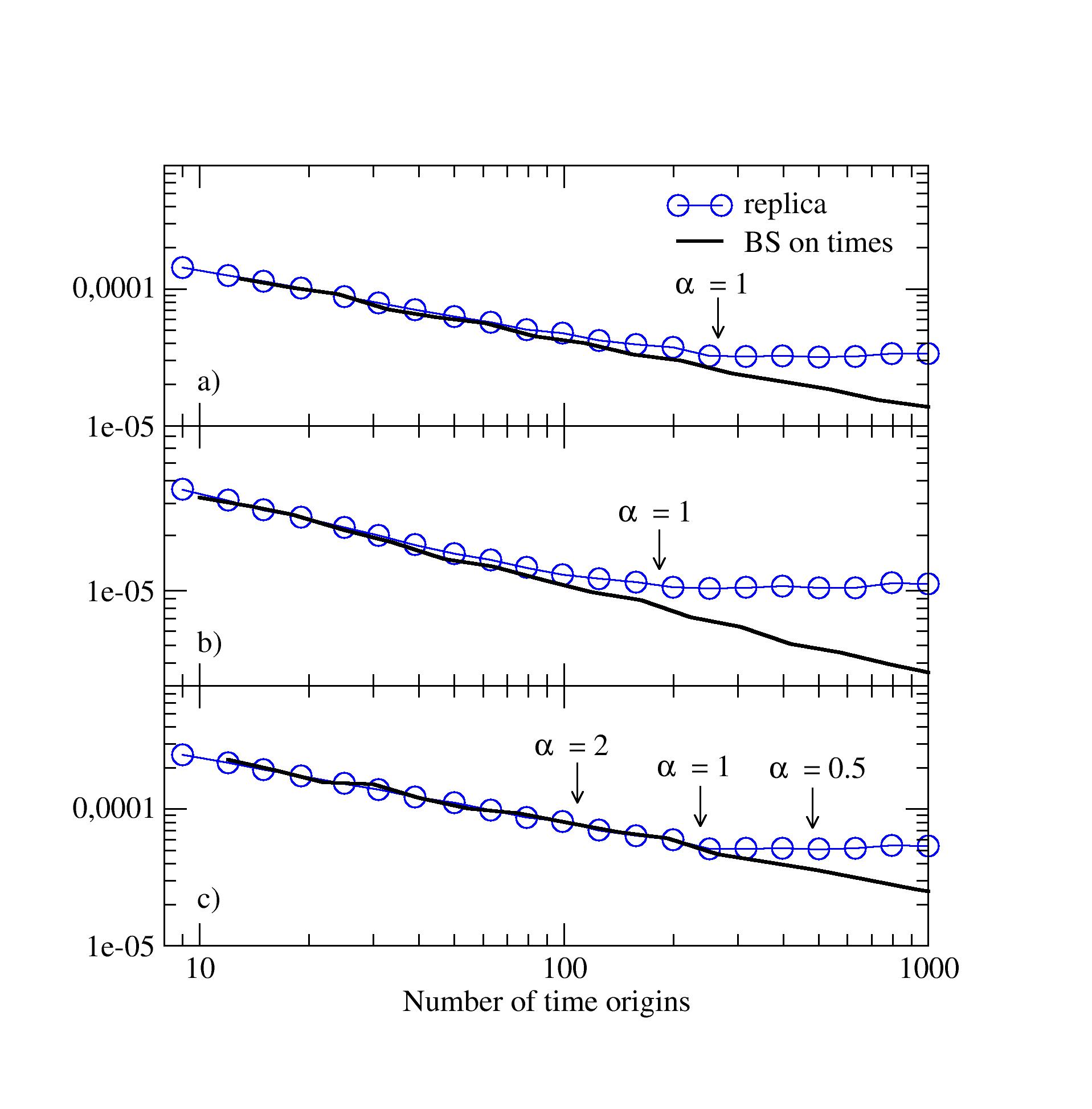}
\caption{Comparison at $\Gamma = 40$ between the BS estimates and the replica results for the mean intrinsic errors on ACF as functions of the number of time origins which depends on the value of $\alpha$: same conventions as in Fig.\,\ref{ACF_04_BS}.}
\label{ACF_40_BS}
\end{center}
\end{figure}

We compare in Fig.\,\ref{ACF_04_BS} to \ref{ACF_40_BS} the BS estimates with the replica results for the mean errors on the VACF of H and He, and on the JACF of the mixture as functions of the number of time origins, corresponding to varying values of $\alpha$, for the three $\Gamma$ values of 0.4, 4, and 40.

For the low numbers of time origins, corresponding to values of $\alpha$ greater than 1, the errors of the methods of BS and replica are comparable. They decrease with increasing numbers of origins, corresponding to decreasing values of $\alpha$, until $\alpha$ reaches unity, where a bifurcation occurs between the BS results and the replica ones. 

For higher numbers of origins, corresponding to values of $\alpha$ lower than 1, the replica results plateau, whereas the BS estimates continue to decrease.

\clearpage

\noindent This behavior originates from the inherent neglect of correlation between time origins in the BS approach, leading to a scaling of the BS error close to $1/\sqrt{M}$, where $M$ is the number of time origins. This scaling is clearly apparent in Fig.\,\ref{ACF_4_BS} and \ref{ACF_40_BS}. In the replica approach, the correlation between successive time origins, that are closer than a correlation time $\tau$,  does not grant more information when comparing several replica.

This comparison between the methods of BS and replica highlights the need to control the number of time origins used in the BS approach. As expected, the origins must be separated by at least one correlation time $\tau$. This leads to the prescription of the value of the control parameter $\alpha = 1$. With this prescription, the BS method provides reliable estimates of the intrinsic error of the ACFs from only one simulation, avoiding to perform several replica simulations.

\begin{figure}[!t]
\begin{center}
\includegraphics[width=.5\textwidth]{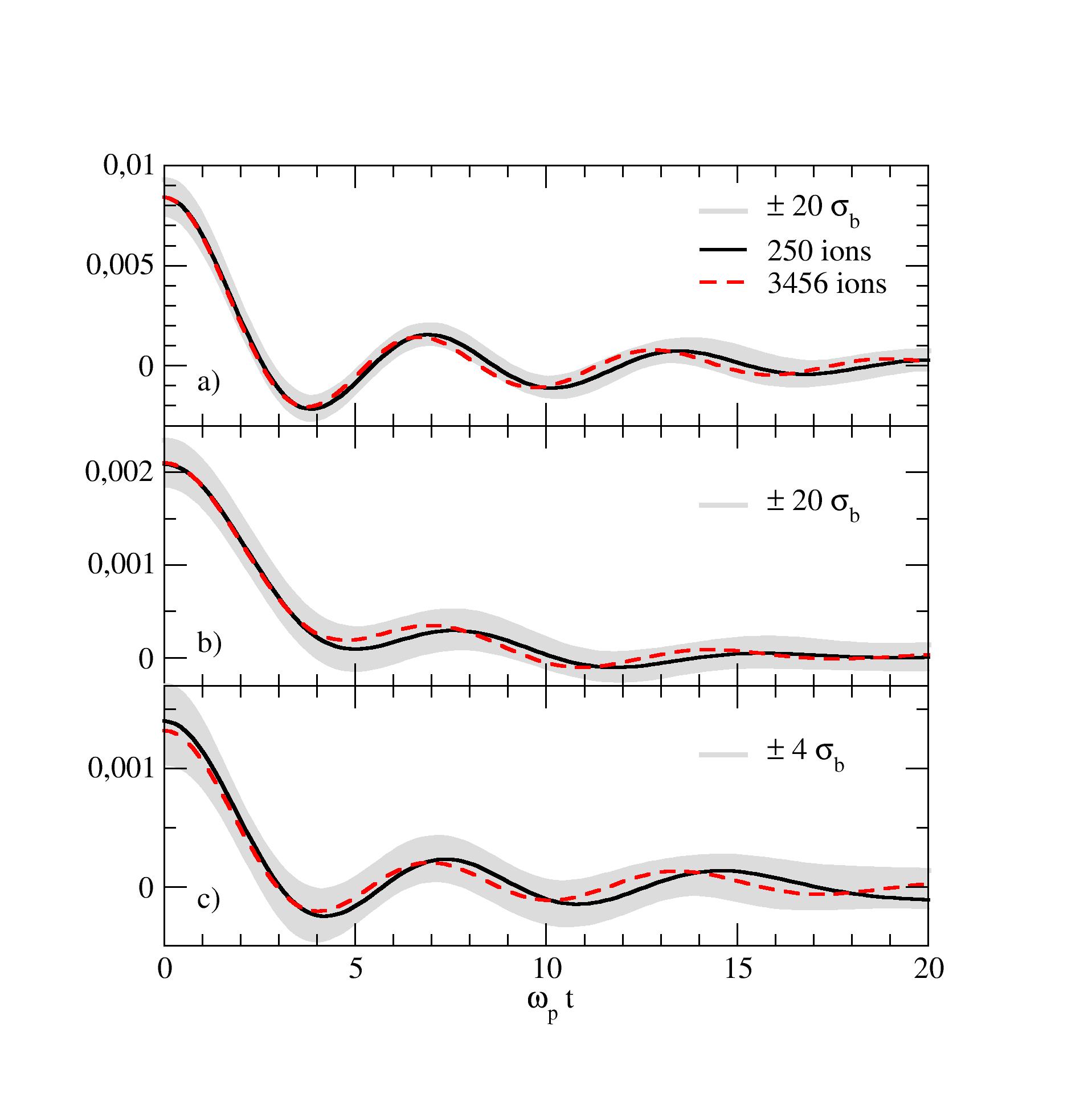}
\caption{Comparison at $\Gamma = 40$ between the ACFs from the simulations using 250 (solid black lines) and 3456 atoms (dashed red lines) as functions of time lag $t$, in unit of inverse plasma frequency $\omega_P$: a) VACF of H, b) VACF of He, c) JACF. The error bars around the ACFs of the simulation with 250 atoms represent a multiple of the standard error $\sigma_b$ estimated bootstrapping both the time origins, with $\alpha = 1$, and the atoms.}
\label{ACF_40_BS_N}
\end{center}
\end{figure}

\begin{figure}[!t]
\begin{center}
\includegraphics[width=.5\textwidth]{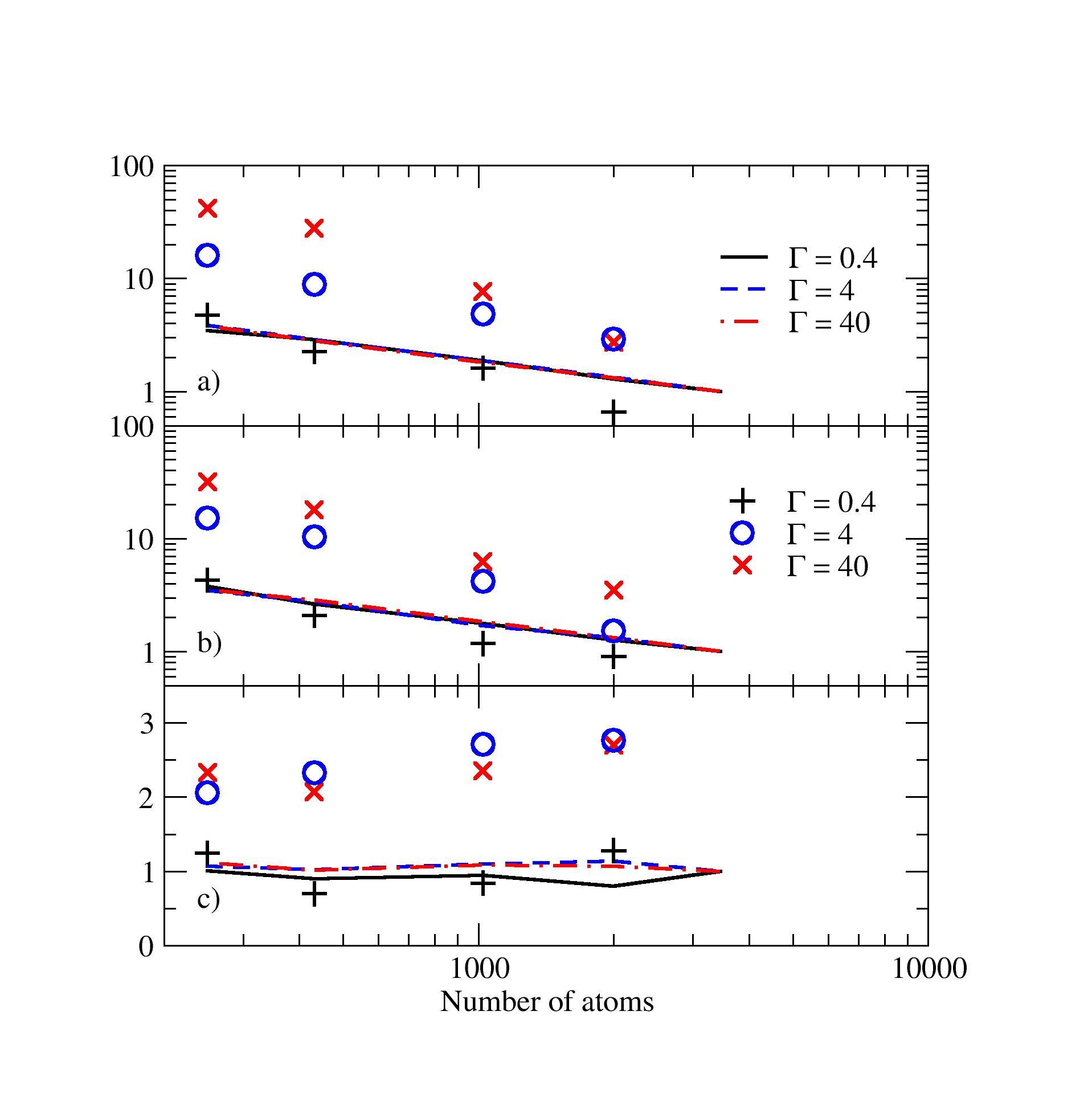}
\caption{Convergence of the ACFs as the number of atoms increases:  a) VACF of H, b) VACF of He, c) JACF. The symbols represent the maximum absolute deviations between the ACF from the simulation at the given number of atoms and the one from the simulation at 3456 atoms: black pluses correspond to $\Gamma = 0.4$, blue open  circles $\Gamma = 4$, red crosses $\Gamma = 40$. The lines represent the mean errors  $\sigma_b$ estimated bootstrapping both the time origins, with $\alpha = 1$, and the atoms: black solid lines correspond to $\Gamma = 0.4$, blue dashed  lines $\Gamma = 4$, red dashed dotted lines $\Gamma = 40$. For each ACF, all values are normalized to the value of $\sigma_b$ corresponding to 3456 atoms.}
\label{ACF_N_cvce}
\end{center}
\end{figure}
\subsection{BS on time origins and atoms}
\label{BSatoms}

We now add to the BS on time origins a BS on atoms to get error estimates covering all the facets of phase space sampling. As a result, the error bars increase by a factor around 2 (see Table\,\ref{table2}). This provides a useful metrics to examine the finite size effects and on how far from the thermodynamic limit a simulation is.

The case with the strongest finite-size effects is at the coupling of $\Gamma = 40$ as can be seen in Fig.\,\ref{ACF_40_BS_N}, where the ACFs computed from a simulation with 250 atoms are compared with the ones computed with 3456 atoms. Error bars are drawn around the results corresponding to 250 atoms as a multiple of the standard errors $\sigma_b$ obtained bootstrapping the time origins (with $\alpha = 1$) and the atoms. The deviations of the VACFs of H and He between 250 and 3456 atoms are large reaching 20 $\sigma_b$, whereas the deviation of the JACF stay within 3 $\sigma_b$. 

It is instructive to describe a synthesis of the convergence of the results with the number of atoms, as displayed in Fig.\,\ref{ACF_N_cvce}. The maximum deviation of an ACF, obtained from a simulation with $N$ atoms, from its values when $N = 3456$ is compared with the mean error $\sigma_b$ determined by a bootstrap of time origins (with $\alpha = 1$) and of the atoms. For each ACF, all the data for different $N$ are normalized to the corresponding value of $\sigma_b$ at $N = 3456$. 

The results of Fig.\,\ref{ACF_40_BS_N} are reported as red crosses symbols and red dashed dotted lines at $N = 250$. The maximum deviations between the VACF of H and He  reach more than 10\,$\sigma_b$, whereas it is only around 2\,$\sigma_b$ for the JACF. Actually, the inspection of Fig.\,\ref{ACF_40_BS_N}\,c) shows that the deviations for the JACF are often less than this latter factor.

In the case of the JACF of the mixture, the results obtained at $N = 250$ do not vary when the number of atoms increases. The convergence with $N$ is reached already at $N = 250$. The interest in the BS determination of $\sigma_b$ is to provide a metrics that can relate the results of only two simulations with increased number of atoms. Comparing $N = 250$ with $N = 432$, for instance, indicates that there is little to expect by increasing further the number of atoms. Any attempt to improve the accuracy of the JACF must pass by an increase in the duration of the simulation or by the production of replica.

In the cases of the VACF of H and He, both the maximum deviations and the BS mean error $\sigma_b$ decrease as the number of atoms $N$ increases. The decrease of $\sigma_b$ is close to the scaling of $1/\sqrt{N}$ expected for the ACF of an individual observable that is averaged over the number of atoms. It was already observed in Table\,\ref{table2} when comparing the intrinsic errors of the VACF and the JACF. 

At the lowest coupling of $\Gamma = 0.4$, the maximum deviations are close to $\sigma_b$ suggesting that there is no finite size effects. Gains in accuracy can be obtained by increasing the number of atoms, but also the duration of the simulation or the production of replica.

At the strong couplings of $\Gamma$\,=\,4 and 40, the ratio between the maximum deviation and $\sigma_b$ also decreases from around 10 at $N = 250$ to around 2 at $N = 2000$. This suggests that finite size effects vanish at $N = 2000$. When comparing only two simulations with increased number of atoms, the discrepancies between both VACFs can be compared with the values of BS error estimates $\sigma_b$. It is large when comparing simulations at $N = 250$ and 432, and comparable for the simulations at $N = 1000$ and 2000. With such kind of comparisons, the BS method can assist in assessing the convergence of the ACFs with the number of atoms.

\subsection{Darken relation revisited}
\label{DarkenRevisited}

As an illustration of the BS method in a typical application, we reexamine the accuracy of the Darken approximation giving the JACF of the mixture as a function of the VACFs of the components. This comparison was performed by Hansen \textit{et al.} \cite{Hansen1985} without error bars. The BS method provides such error quantification, which informs about the quality of the approximation, but also how improvements on the conditions of simulation can lower the errors so as to better constrain the comparison.

The Darken relation, at the level of the ACFs, reads \cite{Hansen1985}
\begin{equation}
\tilde C_{\bold J}(t) = \dfrac{ C_{\bold J}(t)}{ C_{\bold J}(0)} \approx x_2 m_2 \,\tilde C_{\bold v^{(1)}}(t) + x_1 m_1 \,\tilde C_{\bold v^{(2)}}(t),
\end{equation}
with $m_i = M_i/(x_1 M_1+x_2 M_2)$.

\begin{figure}[!t]
\begin{center}
\includegraphics[width=.48\textwidth]{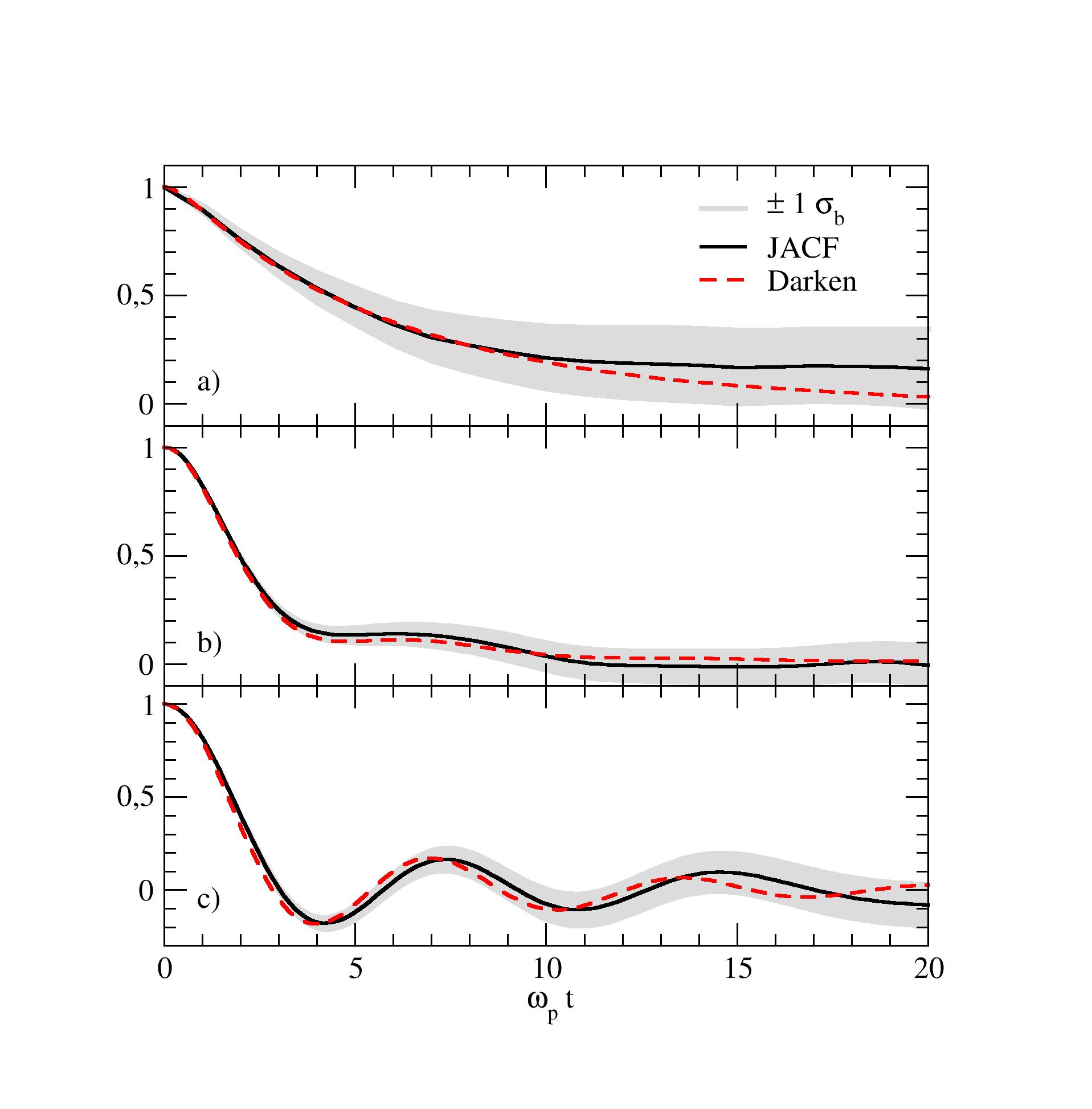}
\caption{Comparison between the JACF (black solid lines), normalized to its value at vanishing time lag, and the Darken approximation (red dashed lines) as functions of time lag $t$, in unit of inverse plasma frequency $\omega_P$, at: a) $\Gamma = 0.4$, b) $\Gamma = 4$, c) $\Gamma = 40$. The error bars around the JACFs represent one standard error $\sigma_b$ estimated bootstrapping both the time origins, with $\alpha = 1$, and the atoms.}
\label{ACF_BS_Darken}
\end{center}
\end{figure}

Since the VACFs are much more accurately determined than the JACF, we decided to neglect the intrinsic error of the Darken expression. In cases where the expressions to compare are determined with comparable accuracy, the BS method can be easily applied to any function of the ACFs, naturally propagating the errors. The Darken expression is compared in Fig.\,\ref{ACF_BS_Darken} to the JACF with error bars obtained by bootstrapping both its time origins (with $\alpha = 1$) and the atoms.

As predicted by Zwanzig \cite{Zwanzig1969} and recalled in Sec.\,\ref{Zwanzig}, the intrinsic error on the normalized JACF are greatly reduced at short time lag. It suffices to compare Fig.\,\ref{ACF_BS_Darken} with panels c) of Fig.\,\ref{ACF_04_replica} to \ref{ACF_40_replica}.

The Darken approximation is within 1-2\,$\sigma$ from the JACF, confirming its pertinence, but at the weakest coupling of $\Gamma = 0.4$ the error bars are much larger than at stronger coupling. There is room for improvement in this low coupling case to reduce the error bars by performing longer simulation and/or additional replica, as was discussed in the preceding sections.

The finite size effects must also be examined since the VACFs are more sensitive to them than the JACF (see Sec.\,\ref{BSatoms}). We checked that the picture obtained with $N = 250$, does not change when increasing $N$.

\section{Conclusion}
\label{conclusion}

The bootstrap (BS) method is a promising approach for the quantification of uncertainty pertaining to the determination by molecular dynamics of time correlation functions. The point is that the usual methods involve long duration of the simulation and/or production of a large number of replica simulations to perform a statistical analysis by blocks \cite{Allen2017}. At contrast, the BS method represents a tremendous shortcut since it relies on only one simulation to produce error estimates. This direct uncertainty quantification can lead to important decisions on how to improve the accuracy through longer simulation or a larger number of atoms.

We have determined the conditions of application of the BS method for the autocorrelation functions (ACF) of individual and collective observables: 
\begin{itemize}
\item The time sequences, that are averaged under the ergodic assumption, must be separated by a time scale of correlation to avoid redundancies. A comparison with the replica method indicates that Zwanzig's correlation time defines an appropriate time scale \cite{Zwanzig1969}.
\item When bootstrapping atoms in the expression of a collective dynamic variable, the BS method is no longer an estimator of the mean and we have demonstrated that a renormalization of each BS sampling is necessary.
\end{itemize}

These findings were confirmed by the analysis of molecular simulations of a binary mixture of ions in Coulomb interaction \cite{Hansen1985}. As an individual observable, we chose the particle velocity, and as a collective one, the interdiffusion current. Their ACFs provide the transport coefficients of self-diffusion and mutual diffusion \cite{Hansen2006}, that are important ingredients of the modeling of mixing layers \cite{Mackay2020}. This example of application is instructive and pertinent since the BS method allowed us to highlight every pieces of uncertainty affecting the powerful Darken approximation \cite{Darken1948,Liu2011} relating the interdiffusion to the self-diffusions of each component of the mixture. We have addressed the issues of phase space sampling and finite size effects within this framework. The monitoring of the convergence to the thermodynamic limit as the number of atoms increases is largely eased with error bars obtained from only one simulation via the BS method.

More applications of the BS method to time correlation functions of molecular dynamics would be welcome to study it in other contexts: shorter range potentials, different observables, \textit{ab initio} simulations... 

In a second step of uncertainties quantification, we intend to use the BS error estimates on ACFs to better control the determination of transport coefficients within a Bayesian inference of the parameters of analytic fitting functions. 

\section{Acknowledgements}

Luc Kazandjian is acknowledged for fruitful discussions.



\section{References}

%

 \appendix
 
\section{OCP system of units}
\label{appOCP}
 
The OCP units of space and time are respectively the Wigner-Seitz radius $a$ and the inverse plasma frequency $1/\omega_P$. Let us note $\tilde r = r/a$ and $\tilde t = \omega_P t$ the reduced variables of space and time. Newton's equations of motion
\begin{subequations}
\begin{equation}
M \, \dfrac{d^2 \bold r_i}{dt^2} = - \sum_{j \ne i} \, \dfrac{Q^2}{r_{ij}^2} ~ \dfrac{\bold r_{ij}}{r_{ij}},
\end{equation}
read in reduced variables
\begin{equation}
\dfrac{d^2 \tilde {\bold r}_i}
{d \tilde t^2} = - \frac{1}{3} \sum_{j \ne i} \, \dfrac{1}{\tilde r_{ij}^2} ~ \dfrac{\tilde {\bold r}_{ij}}{\tilde r_{ij}},
\end{equation}
\end{subequations}
since
\begin{equation}
\omega_P^2 = \dfrac{4 \pi n \, Q^2}{M},
\end{equation}
and
\begin{equation}
\frac{4 \pi}{3} a^3 ~ n = 1,
\end{equation}
where $M$ is the particle mass and $Q$ its charge.

As a consequence, there is no parameter in the equations of motion in reduced variables. An alternative view is to consider that the mass $M = 1$ is constant, as well as the charge $Q = 1/\sqrt{3}$. Moreover, the density is constant equal to 
\begin{equation}
\tilde n = \frac{3}{4 \pi}.
\end{equation}
The only free parameter is the reduced temperature $\tilde T$ that is given from $T$ by the following calculation of the mean kinetic energy $K$ per particle
\begin{subequations}
\begin{equation}
K = \frac{1}{N} \sum_i \frac{1}{2}\,M\,\left(\dfrac{d \bold r_i}{dt}\right)^2 = \frac{3}{2}\,k_B T,
\end{equation}
which reads in reduced variables
\begin{equation}
\tilde K = \frac{1}{N} \sum_i \frac{1}{2}\,\left(\dfrac{d \tilde{\bold r}_i}{d\tilde t}\right)^2 = \frac{3}{2}\,k_B \tilde T,
\end{equation}
with
\begin{equation}
k_B \tilde T = \dfrac{1}{3 \, \Gamma},
\end{equation}
where $\Gamma = Q^2 / a k_B T$ is the Coulomb coupling parameter.
\end{subequations}

It is therefore evidence that every static or dynamic properties of the OCP depend only on the coupling parameter $\Gamma$ (through the reduced temperature) when these properties are expressed in OCP units.

\section{Algorithm for NVK dynamics}
\label{appNVK}
 
 The velocity Verlet algorithm has the advantage to start from a sole iteration where the positions $\bold r_i$, the velocities $\bold v_i$, and the forces $\bold F_i$ acting on each particle, of mass $M_i$, are simultaneously given. It reads
\begin{subequations}
 \begin{equation}
 \bold r_i^{(n+1)} = \bold r_i^{(n)} + \bold v_i^{(n)}\,\Delta + \dfrac{\bold F_i^{(n)}}{2 M_i}\,\Delta^2,
 \end{equation}
 \begin{equation}
 \bold F_i^{(n+1)} = \bold F(\bold r_1^{(n+1)}, \dots , \bold r_N^{(n+1)}),
 \end{equation}
 \begin{equation}
 \bold v_i^{(n+1)} = \bold v_i^{(n)} +\dfrac{\bold F_i^{(n)} + \bold F_i^{(n+1)}}{2 M_i}\,\Delta,
 \end{equation}
\end{subequations}
where $\Delta$ is the time step, and the value of a quantity $X$ at time $t = n \Delta$ is denoted $X^{(n)} = X(n \Delta)$. At the next time step, one computes first the new positions, second the new forces, and third the new velocities.

In the NVK ensemble, the trajectories of the atoms do not follow Newton's equations of motion but the followings
\begin{subequations}
\begin{equation}
M_i \dfrac{d \bold v_i}{dt} = \bold F_i - \alpha M_i \bold v_i,
\end{equation}
with
\begin{equation}
\alpha = \dfrac{1}{2 K} \sum_{j=1}^N \bold F_j\cdot \bold v_j,
\end{equation}
where
\begin{equation}
K = \frac{1}{2} \sum_{j=1}^N M_j v_j^2.
\end{equation}
\end{subequations}
The velocity Verlet algorithm then reads
\begin{subequations}
 \begin{equation}
 \bold r_i^{(n+1)} = \bold r_i^{(n)} + \bold v_i^{(n)}\,\Delta + \dfrac{\bold F_i^{(n)}}{2 M_i}\,\Delta^2 -\frac{1}{2} \alpha^{(n)} \bold v_i^{(n)}\,\Delta^2,
 \end{equation}
 \begin{equation}
 \bold F_i^{(n+1)} = \bold F(\bold r_1^{(n+1)}, \dots , \bold r_N^{(n+1)}),
 \end{equation}
 \begin{align}
 \bold v_i^{(n+1)} = & ~\bold v_i^{(n)} +\dfrac{\bold F_i^{(n)} + \bold F_i^{(n+1)}}{2 M_i}\,\Delta \\
 &- \frac{1}{2} \left[\alpha^{(n)} \bold v_i^{(n)} +  \alpha^{(n+1)} \bold v_i^{(n+1)}\right]\,\Delta. \nonumber
 \end{align}
\end{subequations}
Although $\alpha^{(n+1)}$ can be Taylor-expanded around $\bold v_i^{(n)}$ at first order in $\Delta$, we made the following approximation
\begin{equation}
 \alpha^{(n+1)} = \alpha^{(n)},
 \end{equation} 
 leading to a simpler expression of $\bold v_i^{(n+1)}$.

\end{document}